# OPTIMAL REQUIREMENTS

# OF A DATA ACQUISITION SYSTEM

# FOR A QUADRUPOLAR PROBE

# EMPLOYED IN ELECTRICAL SPECTROSCOPY

A. Settimi*, A. Zirizzotti, J. A. Baskaradas, C. Bianchi

INGV (Istituto Nazionale di Geofisica e Vulcanologia) –

via di Vigna Murata 605, I-00143 Rome, Italy

*Corresponding author: Dr. Alessandro Settimi

Istituto Nazionale di Geofisica e Vulcanologia (INGV)

Via di Vigna Murata 605

I-00143 Rome, Italy

Tel: +39-0651860719

Fax: +39-0651860397

Email: alessandro.settimi@ingv.it

# Abstract

This paper discusses the development and engineering of electrical spectroscopy for simultaneous and non invasive measurement of electrical resistivity and dielectric permittivity. A suitable quadrupolar probe is able to perform measurements on a subsurface with inaccuracies below a fixed limit (*10%*) in a bandwidth of low (LF) frequency (*100kHz*). The quadrupole probe should be connected to an appropriate analogical digital converter (ADC) which samples in phase and quadrature (IQ) or in uniform mode. If the quadrupole is characterized by a galvanic contact with the surface, the inaccuracies in the measurement of resistivity and permittivity, due to the IQ or uniform sampling ADC, are analytically expressed. A large number of numerical simulations proves that the performances of the probe depend on the selected sampler and that the IQ is better compared to the uniform mode under the same operating conditions, i.e. bit resolution and medium.

# 1. Introduction.

*Analogical to Digital Converter (ADC)(Razavi, 1995).* Typically, an ADC is an electronic device that converts an input analogical voltage (or current) to a digital number.

A sampler has several sources of errors. Quantization error and (assuming the sampling is intended to be linear) non-linearity is intrinsic to any analog-to-digital conversion. There is also a so-called aperture error which is due to clock jitter and is revealed when digitizing a time-variant signal (not a constant value). The accuracy is mainly limited by quantization error. However, a faithful reproduction is only possible if the sampling rate is higher than twice the highest frequency of the signal. This is essentially what is embodied in the Shannon-Nyquist sampling theorem.

There are currently a huge number of papers published in scientific literature, and the multifaceted nature of each one makes it difficult to present a complete overview of the ADC models available today. Technological progress, which is rapidly accelerating, makes this task even harder. Clearly, models of advanced digitizers must match the latest technological characteristics.

Different users of sampler models are interested in different modelling details, and so numerous models are proposed in scientific literature: some of them describe specific error sources (Polge et al., 1975); others are devised to connect conversion techniques and corresponding errors (Arpaia et al., 1999)(Arpaia et al., 2003); others again are devoted to measuring the effect of each error source in order to compensate it (Björsell and Händel, 2008). Finally, many papers (Kuffel et al., 1991)(Zhang and Ovaska, 1998) suggest general guidelines for different models.

*Electrical spectroscopy.* Electrical resistivity and dielectric permittivity are two independent physical properties which characterize the behavior of bodies when these are excited by an electromagnetic field. The measurements of these properties provides crucial information regarding practical uses of bodies (for example, materials that conduct electricity) and for countless other purposes.

Some papers (Grard, 1990a,b)(Grard and Tabbagh, 1991)(Tabbagh et al., 1993)(Vannaroni et al. 2004)(Del Vento and Vannaroni, 2005) have proved that electrical resistivity and dielectric permittivity can be obtained by measuring complex impedance, using a system with four electrodes, but without requiring resistive contact between the electrodes and the investigated body. In this case, the current is made to circulate in the body by electric coupling, supplying the electrodes with an alternating electrical signal of low (LF) or middle (MF) frequency. In this type of investigation the range of optimal frequencies for electrical resistivity values of the more common materials is between *≈10kHz* and *≈1MHz*. Once complex impedance has been acquired, the distributions of resistivity and permittivity in the investigated body are estimated using well-known algorithms of inversion techniques.

Applying the same principle, but limited to the acquisition only of resistivity, there are various commercial instruments used in geology for investigating the first 2-5 meters underground both for the exploration of environmental areas and archaeological investigation (Samouëlian et al., 2005).

As regards the direct determination of the dielectric permittivity in subsoil, omitting geo-radar which provides an estimate by complex measurement procedures on radar-gram processing (Declerk, 1995)(Sbartaï et al., 2006), the only technical instrument currently used is the so-called time-domain reflectometer (TDR), which utilizes two electrodes inserted deep in the ground in order to acquire this parameter for further analysis (Mojid et al., 2003)(Mojid and Cho, 2004).

### 1.1. Topic and structure of the paper.

This paper presents a discussion of theoretical modelling and moves towards a practical implementation of a quadrupolar probe which acquires complex impedance in the field, filling the technological gap noted above.

A quadrupolar probe allows measurement of electrical resistivity and dielectric permittivity using alternating current at LF ($30kHz<f<300kHz$) or MF ($300kHz<f<3MHz$) frequencies. By increasing the distance between the electrodes, it is possible to investigate the electrical properties of sub-surface structures to greater depth. In appropriate arrangements, measurements can be carried out with the electrodes slightly raised above the surface, enabling completely non-destructive analysis, although with greater error. The probe can perform immediate measurements on materials with high resistivity and permittivity, without subsequent stages of data analysis.

The authors' paper (Settimi et al., 2010) proposed a theoretical modelling of the simultaneous and non invasive measurement of electrical resistivity and dielectric permittivity, using a quadrupole probe on a subjacent medium. A mathematical-physical model was applied on propagation of errors in the measurement of resistivity and permittivity based on the sensitivity functions tool (Murray-Smith, 1987). The findings were also compared to the results of the classical method of analysis in the frequency domain, which is useful for determining the behaviour of zero and pole frequencies in the linear time invariant (LTI) circuit of the quadrupole. The authors underlined that average values of electrical resistivity and dielectric permittivity may be used to estimate the complex impedance over various terrains (Edwards, 1998) and concretes (Polder et al., 2000)(Laurents, 2005), especially when they are characterized by low levels of water saturation or content (Knight and Nur, 1987) and analyzed within a frequency bandwidth ranging only from LF to MF frequencies (Myounghak et al., 2007)(Al-Qadi et al., 1995). In order to meet the design specifications which ensure satisfactory performances of the probe (inaccuracy no more than *10%*), the forecasts provided by the sensitivity functions approach are less stringent than those foreseen by the transfer functions method (in terms of both a larger band of frequency $f$ and a wider measurable range of resistivity $\rho$ or permittivity $\varepsilon_r$) [see references therein (Settimi et al, 2010)] .

This paper discusses the development and engineering of electrical spectroscopy for simultaneous and non invasive measurement of electrical resistivity and dielectric permittivity. A suitable quadrupolar probe is able to perform measurements on a subsurface with inaccuracies below a fixed limit (*10%*) in a bandwidth of LF (*100kHz*). The quadrupole probe should be connected to an appropriate analogical digital converter (ADC) which samples in phase and quadrature (IQ) (Jankovic and Öhman, 2001) or in uniform mode. If the quadrupole is characterized by a galvanic contact with the surface, the inaccuracies in the measurement of the resistivity and permittivity, due to the IQ or uniform sampling ADC, are analytically expressed. A large number of numerical simulations proves that the performances of the probe depend on the selected sampler and that the IQ is better compared to the uniform mode under the same operating conditions, i.e. bit resolution and medium. Assuming that the electric current injected in materials and so the voltage measured by probe are quasi-monochromatic signals, i.e. with a very narrow frequency band, an IQ down-sampling process can be employed (Oppenheim et al., 1999). Besides the quantization error of IQ ADC, which can be assumed small both in amplitude and phase, as decreasing exponentially with the bit resolution, the electric signals are affected by two additional noises. The amplitude term noise, due to external environment, is modeled by the signal to noise ratio which can be reduced performing averages over a thousand of repeated measurements. The phase term noise, due to a phase-splitter detector, which, even if increasing linearly with the frequency, can be minimized by digital electronics providing a rise time of few nanoseconds. Instead, in order to analyze the complex impedance measured by the quadrupole in Fourier domain, an uniform ADC, which is characterized by a sensible phase inaccuracy depending on frequency, must be connected to a Fast Fourier Transform (FFT) processor, that is especially affected by a round-off amplitude noise linked to both the FFT register length and samples number (Oppenheim et al., 1999). If the register length is equal to *32* bits, then the round-off noise is entirely negligible, else, once bits are reduced to *16*, a technique of compensation must occur. In fact, oversampling can be employed within a short time window, reaching a compromise between the needs of limiting the phase inaccuracy due to ADC and not raising too much the number of averaged FFT values sufficient to bound the round-off.

The paper is organized as follows. Section 2 defines the data acquisition system. In sec. 3, the theoretical modeling is provided for both IQ (3.1) and uniform (3.2) samplers. In sec. 4, assuming quasi-monochromatic signals, an IQ down-sampling process is employed. Besides quantization error of IQ ADC, the electric signals are affected by two additional noises: the amplitude term noise, due to external environment; and the phase term noise, due to phase-splitter detector. In sec. 5, in order to analyze the complex impedance measured by the quadrupolar probe in Fourier domain, the uniform sampling ADC is connected to a FFT processor being affected by a round-off noise. In sec. 6, the design of the characteristic geometrical dimensions of the probe is analyzed. Sec. 7 proposes a conclusive discussion. Finally, an Appendix presents an outline of the somewhat lengthy calculations.

## 2. Data acquisition system.

In order to design a quadrupole probe [fig. 1.a] which measures the electrical conductivity $\sigma$ and the dielectric permittivity $\varepsilon_r$ of a subjacent medium with inaccuracies below a prefixed limit (*10%*) in a band of low frequencies ($B=100kHz$), the probe can be connected to an appropriate analogical digital converter (ADC) which performs a uniform or in phase and quadrature (IQ) sampling (Razavi, 1995)(Jankovic and Öhman, 2001), with bit resolution not exceeding *12*, thereby rendering the system of measurement (voltage scale of $4V$) almost insensitive to the electric noise of the external environment ($\approx 1mV$).

IQ can be implemented using a technique that is easier to realize than in the uniform mode, because the voltage signal of the probe is in the frequency band of $B=100kHz$ and IQ sampled with a rate of only $f_S=4B=400kHz$, while, for example, low resolution uniform samplers are specified for rates of *5-200MHz*.

With the aim of investigating the physics of the measuring system, the inaccuracies in the transfer impedance measured by the quadrupolar probe [fig. 1.b], due to uniform or IQ sampling ADCs [fig. 2], are provided.

If, in the stage downstream of the quadrupole , the electrical voltage $V$ is amplified $V_V=A_V \cdot V$ and the intensity of current $I$ is transformed by a trans-resistance amplifier $V_I=A_R \cdot I$, the signals having been processed by the sampler, then:

the inaccuracy $\Delta|Z|/|Z|$ for the modulus of the transfer impedance results from the negligible contributes $\Delta A_V/A_V$ and $\Delta A_R/A_R$, respectively for the voltage and trans-resistance amplifiers, and the predominant one $\Delta|V_V|/|V_V|$ for the modulus of the voltage, due to the sampling,

$$\frac{\Delta|Z|}{|Z|}=\frac{\Delta A_V}{A_V}+\frac{\Delta A_R}{A_R}+2\frac{\Delta|V_V|}{|V_V|}\cong 2\frac{\Delta|V_V|}{|V_V|}, \tag{2.1}$$

the inaccuracies for the modulus of the voltage and the current intensity being equal, $\Delta|V_V|/|V_V|= \Delta|V_I|/|V_I|$;

instead, the inaccuracy $\Delta\Phi_Z/\Phi_Z$ for the initial phase of the transfer impedance coincides with the one $\Delta\varphi_V/\varphi_V$ for the phase of the voltage, due to the sampler,

$$\frac{\Delta\Phi_Z}{\Phi_Z}=\frac{\Delta\varphi_V}{\varphi_V}, \tag{2.2}$$

the initial phase of the current being null, $\varphi_I=0$.

## 3. Theoretical modeling.

As concerns an IQ mode (Jankovic and Öhman, 2001), in which a quartz oscillates with a high figure of merit $Q=10^4$-$10^6$, the inaccuracy $\Delta|Z|/|Z|_{IQ}(n,\varphi_V)$ depends strongly on bit resolution $n$, decreasing as the exponential function $2^{-n}$ of $n$, and weakly on the initial phase of voltage $\varphi_V$, such that [figs. 3.a, a.bis]

$$\left.\frac{\Delta|Z|}{|Z|}\right|_{IQ}\simeq\frac{1}{2^n}[1+\frac{2\pi}{Q}\tan\varphi_V(1+\cos 2\varphi_V)]=\begin{cases}\frac{1}{2^n}(1+\frac{2\pi}{Q}) & , \quad \varphi_V=\varphi_V^{\max}=\frac{\pi}{4}\\ \frac{1}{2^n}(1-\frac{2\pi}{Q}) & , \quad \varphi_V=\varphi_V^{\min}=\pi-\varphi_V^{\max}\end{cases}, \tag{3.1}$$

$\Delta|Z|/|Z|_{IQ}(n,\varphi_V)$ being

$$\lim_{\varphi_V \to \varphi_V^{\lim}} \left. \frac{\Delta|Z|}{|Z|} \right|_{IQ} = \frac{1}{2^n} \tag{3.2}$$

in the limit value

$$\varphi_V^{\lim} = arctg(\frac{Q}{4\pi}) \approx \frac{\pi}{2} \quad . \tag{3.3}$$

Instead, as concerns uniform sampling (Razavi, 1995), the inaccuracy $\Delta|Z|/|Z|_U(n)$ for $|Z|$ depends only on the bit resolution $n$, decreasing as the exponential function $2^{-n}$ of $n$ [fig. 4.a],

$$\left. \frac{\Delta|Z|}{|Z|} \right|_{U} = \frac{1}{2^n} . \tag{3.4}$$

Consequently, for all ADCs, with bit resolution $n$:

if the probe performs measurements on a medium, then the inaccuracy $\Delta\varepsilon_r/\varepsilon_r(f)$ in the measurement of dielectric permittivity $\varepsilon_r$ is characterized by a minimum limit $\Delta\varepsilon_r/\varepsilon_r|_{min}(\varepsilon_r,n)$, interpretable as the "physical bound" imposed on the inaccuracies of the problem, which depends on both $\varepsilon_r$ and the bit resolution $n$, being directly proportional to the factor $(1+1/\varepsilon_r)$, while decreasing as the exponential function $2^{-n}$ of $n$ [fig. 5.a];

if the probe, with characteristic geometrical dimension $L$, performs measurements on a medium, with conductivity $\sigma$ and permittivity $\varepsilon_r$, working within the cut-off frequency $f_T=f_T(\sigma,\varepsilon_r)=\sigma/(2\pi\varepsilon_0(\varepsilon_r+1))$ (Settimi et al., 2010), then the absolute error $E_{|Z|}(L,\sigma,n)$ in the measurement of the modulus for the transfer impedance $|Z|_N(L,\sigma)$ depends on $\sigma$, $L$ and the bit resolution $n$, the error being inversely proportional to both $\sigma$ and $L$, while decreasing as the exponential function $2^{-n}$ of $n$ [fig. 5.b] [In fact, $Z_N(f,L,\sigma,\varepsilon_r)$ is fully characterized by the high frequency pole $f_T=f_T(\sigma,\varepsilon_r)$, which cancels its denominator: the transfer impedance acts as a low-middle frequency band-pass filter with cut-off $f_T=f_T(\sigma,\varepsilon_r)$, in other words the frequency equalizing Joule and displacement current. As discussed below, average values of $\sigma$ may be used over the band ranging from LF to MF, therefore $|Z|_N(L,\sigma)$ is not function of frequency below $f_T$ (Settimi et al., 2010)].

As concerns IQ mode, with a quartz of high merit figure $Q$, the inaccuracy $\Delta\Phi_Z/\Phi_Z|_{IQ}(n,\varphi_V)$ depends both on the bit resolution $n$, decreasing as the exponential function $2^{-n}$ of $n$, and on the voltage phase $\varphi_V$, such that [fig. 3.b]

$$\left.\frac{\Delta\Phi_Z}{\Phi_Z}\right|_{IQ} \simeq \frac{1}{2^n}\frac{\sin(2\varphi_V)}{2\varphi_V}\left(1+\frac{2\pi}{Q}\tan\varphi_V\right) = \begin{cases} \frac{1}{2^n} & , \quad \varphi_V = \varphi_V^{\max} = 0 \\ 0 & , \quad \varphi_V = \varphi_V^{\min} = \pi \end{cases}, \tag{3.5}$$

$\Delta\Phi_Z/\Phi_Z|_{IQ}(n,\varphi_V)$ being very low in $\varphi_V = \pi/2$, due to the high $Q$ [fig. 3.b.bis],

$$\lim_{\varphi_V \to \pi/2} \left.\frac{\Delta\Phi_Z}{\Phi_Z}\right|_{IQ} = \frac{1}{2^{n-2}}\frac{1}{Q}. \tag{3.6}$$

Instead, as concerns uniform sampling, the inaccuracy $\Delta\Phi_Z/\Phi_Z|_U(f,f_S)$ for $\Phi_Z$ depends on both the working frequency $f$ of the probe and the rate sampling $f_S$ of the ADC, the inaccuracy being directly proportional to the frequency ratio $f/f_S$ [fig. 4.b],

$$\left.\frac{\Delta\Phi_Z}{\Phi_Z}\right|_{U} = 2\frac{f}{f_S}. \tag{3.7}$$

As a consequence, only for uniform sampling ADCs, the inaccuracy $\Delta\Phi_Z/\Phi_Z\ (f,f_S)$ for the phase $\Phi_Z$ must be optimized in the upper frequency $f_{up}$, so when the probe performs measurements at the limit of its band $B$, i.e. $f_{up}=B$.

Still with the aim of investigating the physics of the measuring system, when the quadrupolar probe exhibits a galvanic contact with the subjacent medium of electrical conductivity $\sigma$ and dielectric permittivity $\varepsilon_r$, i.e. $h=0$, and works in frequencies $f$ lower than the cut-off frequency $f_T=f_T(\sigma,\varepsilon_r)$ (Grard and Tabbagh, 1991),

$$\Omega = \frac{\omega}{\omega_T} \le 1, \tag{3.8}$$

the inaccuracies $\Delta\sigma/\sigma$ in the measurements of the conductivity $\sigma$ and $\Delta\varepsilon_r/\varepsilon_r$ for the permittivity $\varepsilon_r$ are analytically expressed [figs. 6, 9, 11][tabs. 1-3], achievable connecting uniform or IQ samplers, which ensure the inaccuracies $\Delta|Z|/|Z|$ for the modulus $|Z|$ and $\Delta\Phi_Z /\Phi_Z$ for the phase $\Phi_Z$ of the transfer impedance (Settimi et al., 2010),

$$\frac{\Delta\sigma}{\sigma} \cong 2(1+\Omega^2)(\frac{\Delta|Z|}{|Z|}+\frac{\Delta\Phi_Z}{\Phi_Z}), \tag{3.9}$$

$$\frac{\Delta\varepsilon_r}{\varepsilon_r} \cong 2(1+\Omega^2)(1+\frac{1}{\varepsilon_r})(\frac{1}{\Omega^2}\frac{\Delta|Z|}{|Z|}+\frac{\Delta\Phi_Z}{\Phi_Z}). \tag{3.10}$$

Only if the quadrupole probe is in galvanic contact with the subjacent medium, i.e. $h=0$, then our mathematical-physical model predicts that the inaccuracies $\Delta\sigma/\sigma$ for $\sigma$ and $\Delta\varepsilon_r/\varepsilon_r$ for $\varepsilon_r$ are invariant in the linear (Wenner's) or square configuration and independent of the characteristic geometrical dimension of the quadrupole, i.e. electrode-electrode distance $L$ (Settimi et al., 2010). If the quadrupole, besides grazing the medium, measures $\sigma$ and $\varepsilon_r$ working in a frequency $f$ much lower than the cut-off frequency $f_T=f_T(\sigma,\varepsilon_r)$, then the inaccuracy $\Delta\sigma/\sigma=F(\Delta|Z|/|Z|,\Delta\Phi_Z/\Phi_Z)$ is a linear combination of the inaccuracies, $\Delta|Z|/|Z|$ and $\Delta\Phi_Z/\Phi_Z$, for the transfer impedance, while the inaccuracy $\Delta\varepsilon_r/\varepsilon_r=F(\Delta|Z|/|Z|)$ can be approximated as a linear function only of the inaccuracy $\Delta|Z|/|Z|$; in other words, if $f<<f_T$, then $\Delta\Phi_Z/\Phi_Z$ is contributing in $\Delta\sigma/\sigma$ but not in $\Delta\varepsilon_r/\varepsilon_r$.

As mentioned above, even if, according to Debye polarization mechanisms (Debye, 1929) or Cole-Cole diagrams (Auty and Cole, 1952), the complex permittivity of various materials in the frequency band from VLF to VHF exhibits several intensive relaxation effects and a non-trivial dependence on the water saturation (Chelidze and Gueguen, 1999)(Chelidze et al., 1999), anyway average values of electrical resistivity and dielectric permittivity may be used to estimate the complex impedance over various terrains and concretes, especially when they are characterized by low levels of water content and analyzed within a frequency bandwidth ranging only from LF to MF.

### 3.1. For in phase and quadrature (IQ) sampling.

As concerns IQ sampling ADCs (merit figure $Q$)(Jankovic and Öhman, 2001), the inaccuracy $\Delta\varepsilon_r/\varepsilon_r(f,n)$ in the measurement of permittivity $\varepsilon_r$ is characterized by an optimal working frequency $f_{opt,IQ}(f_T)$, close to the cut-off frequency for the modulus of the transfer impedance $f_T = f_T(\sigma,\varepsilon_r)$, i.e.

$$f_{opt,IQ} \simeq f_T(1+\frac{\pi/2}{Q}) \approx f_T, \qquad (3.11)$$

which is tuned in a minimum value of inaccuracy $\Delta\varepsilon_r/\varepsilon_r|_{min,IQ}(\varepsilon_r,n)$, depending both on $\varepsilon_r$ and the specifications of the sampler, in particular only its bit resolution $n$. The inaccuracy is directly proportional to the factor $(1+1/\varepsilon_r)$, while decreasing as the exponential function $2^{-(n-3)}$ of $n$, such that [fig. 5.a]

$$\left.\frac{\Delta\varepsilon_r}{\varepsilon_r}\right|_{\min,IQ} \simeq (1+\frac{1}{\varepsilon_r})\frac{1}{2^{n-3}}(1+\frac{\pi}{Q}). \qquad (3.12)$$

Consequently, if the frequency $f$ of the probe is much lower than the cut-off frequency $f_T(\sigma,\varepsilon_r)$ for the transfer impedance, then the inaccuracy $\Delta\sigma/\sigma(n)$ for $\sigma$ is a constant, and the inaccuracy $\Delta\varepsilon_r/\varepsilon_r(f,n)$ for $\varepsilon_r$ shows a downward trend in frequency, as $(f_T/f)^2$, such that both the inaccuracies decrease as exponential functions of $n$, the first inaccuracy as $1/2^{n-2}$ while the second one as $1/2^{n-1}$, i.e. [fig. 6] [tab. 1]

$$\left.\frac{\Delta\sigma}{\sigma}\right|_{IQ} \simeq \frac{1}{2^{n-2}}(1+\frac{\pi}{Q}) \quad , \quad for \quad f << f_T, \qquad (3.13)$$

$$\left.\frac{\Delta\varepsilon_r}{\varepsilon_r}\right|_{IQ} \simeq (1+\frac{1}{\varepsilon_r})\frac{1}{2^{n-1}}(1+\frac{2\pi}{Q})\left(\frac{f_T}{f}\right)^2 \quad , \quad for \quad f << f_T. \qquad (3.14)$$

Even if the frequency $f$ is much higher than $f_T(\sigma,\varepsilon_r)$, it could be proven that the inaccuracies $\Delta\sigma/\sigma(f,n)$ and $\Delta\varepsilon_r/\varepsilon_r(f,n)$ do not deviate too much from an upward trend in frequency as a parabolic line $(f/f_T)^2$, with a high gradient for the $\sigma$ measurement and a low gradient for the $\varepsilon_r$ measurement, still decreasing as exponential functions of $n$, the first inaccuracy as $1/2^{n-2}$ while the second one as $1/2^{n-1}$, i.e. [fig. 6][tab. 1]

$$\left.\frac{\Delta\sigma}{\sigma}\right|_{IQ} \simeq \frac{1}{2^{n-2}}(1+\frac{\pi}{Q})\left(\frac{f}{f_T}\right)^2 \quad , \quad for \quad f >> f_T , \tag{3.15}$$

$$\left.\frac{\Delta\varepsilon_r}{\varepsilon_r}\right|_{IQ} \simeq (1+\frac{1}{\varepsilon_r})\frac{1}{2^{n-1}}\left(\frac{f}{f_T}\right)^2 \quad , \quad for \quad f >> f_T . \tag{3.16}$$

Only if $f$ is lower than $f_T$, then the measurements could be optimized for $\varepsilon_r$ and $\sigma$, and might require that the inaccuracy $\Delta\varepsilon_r/\varepsilon_r(n,f)$ for $\varepsilon_r$ is below the prefixed limit $\Delta\varepsilon_r/\varepsilon_r|_{fixed}$ ($10\%$) within the frequency band $B=100kHz$, choosing a minimum bit resolution $n_{min,IQ}(f_T,\varepsilon_r)$, which depends on both $f_T$ and $\varepsilon_r$ and increases as the logarithmic function $log_2$ of both the ratios $f_T/B$ and $1/\varepsilon_r$, i.e.

$$n_{\min,IQ} \approx 1-\log_2\left.\frac{\Delta\varepsilon_r}{\varepsilon_r}\right|_{fixed} + 2\log_2\frac{f_T}{B} + \log_2(1+\frac{1}{\varepsilon_r}) . \tag{3.17}$$

Referring to the IQ sampling ADCs, the inaccuracies $\Delta\sigma/\sigma$ in the measurement of the electrical conductivity $\sigma$ and $\Delta\varepsilon_r/\varepsilon_r$ for the dielectric permittivity $\varepsilon_r$ were estimated for the worst case, when the inaccuracies $\Delta|Z|/|Z|_{IQ}(n,\varphi_V)$ in the measurement of the modulus and $\Delta\Phi_Z/\Phi_Z|_{IQ}(n,\varphi_V)$ for the phase of the transfer impedance respectively assume the mean and the maximum values, i.e. $\Delta|Z|/|Z|_{IQ} = \Delta\Phi_Z/\Phi_Z|_{IQ} = 1/2^n$.

To conclude, for the IQ mode, the optimal and minimum values of the working frequency of the quadrupolar probe interact in a competitive way. The more the quadrupole analyzes a subjacent medium characterized by a low electrical conductivity, with the aim of shifting the optimal frequency into a low band, the more the low conductivity has the self-defeating effect of shifting the minimum value of frequency into a higher band. In fact, the probe could work around a low optimal frequency, achievable in measurements of transfer impedance with a low cut-off frequency, typical of materials characterized by low conductivity. Instead, the more the minimum value of the working frequency is shifted into a lower band, the more the minimum bit resolution for the sampler has to be increased; if the medium was selected, then increasing the inaccuracy for the measurement of the dielectric permittivity, or, if the inaccuracy was fixed, shifting the cut-off frequency into a high band, i.e. selecting a medium with high conductivity. Finally, while the authors' analysis shows that the quadrupole could work at a low optimal frequency, if the transfer impedance is characterized by a low cut-off frequency, in any case and in accordance with the more traditional results of recent scientific publications referenced (Grard, 1990, a-b)(Grard and Tabbagh, 1991)(Tabbagh et al., 1993)(Vannaroni et al. 2004)(Del Vento and Vannaroni, 2005), the probe could perform measurements in an appropriate band of higher frequencies, centered around the cut-off frequency, where the inaccuracy for the measurements of conductivity and permittivity were below a prefixed limit (*10%*).

### 3.2. For uniform sampling.

As concerns uniform sampling ADCs (Razavi, 1995), if the frequency $f$ of the probe is lower than the cut-off frequency $f_T$ for the modulus $|Z|(f,L)$ of the transfer impedance, i.e. $f_T = f_T(\sigma,\varepsilon_r)$, then, the higher the bit resolution $n$, the more the optimal working frequency $f_{opt,U}(f_T,f_S,n)$, which minimizes inaccuracy $\Delta\varepsilon_r/\varepsilon_r(f,f_S,n)$ in the measurement of the permittivity $\varepsilon_r$, approximately depends on the cut-off frequency $f_T(\sigma,\varepsilon_r)$ and the specifications of the sampler, in particular the sampling rate $f_S$ and $n$, increasing with both $f_T$ and $f_S$, while decreasing as the exponential function $2^{-n/3}$ of $n$, such that [fig. 7]

$$\frac{f_{opt,U}}{f_T} \simeq \sqrt[3]{\frac{1}{2^n}\frac{f_S}{f_T}} . \tag{3.18}$$

Moreover, the higher the bit resolution $n$, the more the inaccuracy $\Delta\varepsilon_r/\varepsilon_r(f,f_S,n)$ for $\varepsilon_r$ can not go down beyond a minimum limit of inaccuracy $\Delta\varepsilon_r/\varepsilon_r|_{min,U}(\varepsilon_r,n)$, which approximately depends on both $\varepsilon_r$ and $n$, being directly proportional to the factor $(1+1/\varepsilon_r)$, while decreasing as the exponential function $2^{-(n-1)}$ of $n$, similarly to IQ sampling. The minimum value of inaccuracy $\Delta\varepsilon_r/\varepsilon_r|_{min,U}(\varepsilon_r,n)$ in uniform mode is higher than the minimum inaccuracy $\Delta\varepsilon_r/\varepsilon_r|_{min,IQ}(\varepsilon_r,n)$ corresponding to the IQ mode, i.e. [figs. 5.a, 6] [tab. 1]

$$\frac{\Delta\varepsilon_r}{\varepsilon_r} \geq \left.\frac{\Delta\varepsilon_r}{\varepsilon_r}\right|_{\min,U} \simeq (1+\frac{1}{\varepsilon_r})\frac{1}{2^{n-1}}[1+(\frac{f_T}{f_{opt,U}})^2] >> \frac{1}{4}\left.\frac{\Delta\varepsilon_r}{\varepsilon_r}\right|_{\min,IQ} , \quad f_{opt,U} << f_T . \tag{3.19}$$

Finally, the minimum value of frequency $f_{U,min}(f_T,n)$, which allows an inaccuracy $\Delta\varepsilon_r/\varepsilon_r(f,f_S,n)$ below a prefixed limit $\Delta\varepsilon_r/\varepsilon_r|_{fixed}$ (*10%*), depends both on $f_T(\sigma,\varepsilon_r)$ and $n$, being directly proportional to $f_T$, while decreasing as an exponential function of $n$, such that [fig. 8]

$$f_{\min,U} \simeq \frac{f_T}{\sqrt{\frac{\Delta\varepsilon_r/\varepsilon_r|_{fixed}}{\Delta\varepsilon_r/\varepsilon_r|_{\min,U}} - 1}} < f_{opt,U} . \tag{3.20}$$

To conclude, also in the uniform mode, the optimal frequency and the band of frequency of the quadrupolar probe interact in a competitive way. In fact, an ADC with a high bit resolution is characterized by a low sampling rate, for which, having selected the subjacent medium to be analyzed, the higher the resolution of the sampler used, with the aim of shifting the optimal frequency of the quadrupolar probe into a low band, the more the low sampling rate has the self-defeating effect of narrowing its frequency band. Moreover, a material with low electrical conductivity is usually characterized by low dielectric permittivity. Having designed the ADC, the more the quadrupole measures a transfer impedance limited by a low cut-off frequency, the more it can work at a low optimal frequency, even if centered in a narrow band. Finally, having selected the medium to be analyzed and designed the sampler, the more the frequency band of the probe is widened, the more the inaccuracy of the measurements is increased.

# 4. Noisy IQ Down-Sampling.

In signal processing, down-sampling (or "sub-sampling") is the process of reducing the sampling rate of a signal. This is usually done to reduce the data rate or the size of the data (Oppenheim et al., 1999)(Andren and Fakatselis, 1995).

The down-sampling factor, commonly denoted by *M*, is usually an integer or a rational fraction greater than unity. If the quadrupolar probe injects electric current into materials at a RF frequency *f*, then the ADC samples at a rate *fs* fixed by:

$$f_S = \frac{f}{M}, \tag{4.1}$$

being *M* preferably, but not necessary, a power of *2* to facilitate the digital circuitry ($M = 2^m$, $m \in \mathbb{N}$).

Since down-sampling reduces the sampling rate, one must be careful to make sure the Shannon-Nyquist sampling theorem criterion is maintained. If the sampling theorem is not satisfied then the resulting digital signal will have aliasing. To ensure that the sampling theorem is satisfied, a low-pass filter is used as an anti-aliasing filter to reduce the bandwidth of the signal before the signal is down-sampled; the overall process (low-pass filter, then down-sample) is called decimation. Note that the anti-aliasing filter must be a low-pass filter in down-sampling. This is different from sampling a continuous signal, where either a low-pass filter or a band-pass filter may be used.

A practical scheme to select the sampling rate is to launch two time sequences as in fig. 2.a.. Now there is a problem due to timing. If the quadrupole would work at a fixed frequency *f*, then the proper relationship between the rate *fs* and the down-sampling factor *M* could be easily found. Instead, if the probe is performing a sort of electrical spectroscopy, then an enable signal for the sampling and holding circuit (S&H) must be generated. The time frame should be such that the sequences $n{\cdot}M{\cdot}T$ for the sample *I* and $n{\cdot}M{\cdot}T+T/4$ for the sample *Q* could be obtained, corresponding to the period *T* of the maximum working frequency. So the rate *fs* would be ensured as a *M* factor sub-multiple of frequency $f$. A possible conceptual scheme of this implementation is shown in fig. 2.b.

Besides the quantization error of IQ ADC, which can be assumed small both in amplitude and phase, as decreasing exponentially with the bit resolution $n$, the electric signals are affected by two additional noises. The amplitude term noise, due to external environment, is modelled by the signal to noise ratio $SNR = 30dB$ which can be reduced performing averages over one thousand of repeated measurements ($A = 10^3$). The phase term noise, due to a phase-splitter detector, which, even if increasing linearly with the frequency $f$, can be minimized by digital electronics providing a rise time of few nanoseconds ($\tau = 1ns$). In analytical terms:

$$\frac{\Delta|Z|}{|Z|} = \left.\frac{\Delta|Z|}{|Z|}\right|_{IQ} + \frac{1}{\sqrt{A}} \left.\frac{\Delta|Z|}{|Z|}\right|_{Enviroment} = \frac{1}{2^n} + \frac{1}{\sqrt{A}} \frac{2}{SNR}, \quad (4.2)$$

$$\frac{\Delta\Phi_Z}{\Phi_Z} = \left.\frac{\Delta\Phi_Z}{\Phi_Z}\right|_{IQ} + \left.\frac{\Delta\Phi_Z}{\Phi_Z}\right|_{Phase-Shifter} = \frac{1}{2^n} + \tau \cdot f. \quad (4.3)$$

With respect to the ideal case involving only a quantization error, the additional noise both in amplitude, due to external environment, and especially in phase, due to phase-shifter detector, produce two effects: firstly, both the curves of inaccuracy $\Delta\varepsilon_r/\varepsilon_r(f)$ and $\Delta\sigma/\sigma(f)$ in measurement of the dielectric permittivity $\varepsilon_r$ and electric conductivity $\sigma$ are shifted upwards, to values larger of almost half a magnitude order, at most; and, secondly, the inaccuracy curve $\Delta\varepsilon_r/\varepsilon_r(f)$ of permittivity $\varepsilon_r$ is narrowed, even of almost half a middle frequency (MF) decade. So, both the optimal value of frequency $f_{opt}$, which minimizes the inaccuracy $\Delta\varepsilon_r/\varepsilon_r(f)$ of $\varepsilon_r$, and the maximum frequency $f_{max}$, allowing an inaccuracy $\Delta\varepsilon_r/\varepsilon_r(f)$ below the prefixed limit $\Delta\varepsilon_r/\varepsilon_r|_{fixed}$ (*10%*), are left shifted towards lower frequencies, even of half a MF decade. Instead, the phase-splitter is affected by a noise directly proportional to frequency, which is significant just from MFs; so, the minimum frequency $f_{min}$, allowing $\Delta\varepsilon_r/\varepsilon_r(f)$ below $\Delta\varepsilon_r/\varepsilon_r|_{fixed}$ (*10%*), remains almost invariant at LFs [fig. 9][tab. 2].

Therefore, the profit of employing the down-sampling method is obvious. This method would allow to run a real electric spectroscopy because, theoretically, measurements could be performed at any frequency. An advantage is that there are virtually no limitations due to the sampling rate of ADCs and associated S&H circuitry [fig. 10][tab. 2].

## 5. Fast Fourier Transform (FFT) processor and round-off noise.

In mathematics, the Discrete Fourier Transform (DFT) is a specific kind of Fourier transform, used in Fourier analysis. The DFT requires an input function that is discrete and whose non-zero values have a limited (finite) duration. Such inputs are often created by sampling a continuous function. Using the DFT implies that the finite segment that is analyzed is one period of an infinitely extended periodic signal; if this is not actually true, a window function has to be used to reduce the artefacts in the spectrum. In particular, the DFT is widely employed in signal processing and related fields to analyze the frequencies contained in a sampled signal. A key enabling factor for these applications is the fact that the DFT can be computed efficiently in practice using a Fast Fourier Transform (FFT) algorithm.

It is important to understand the effects of finite register length in the computation. Specifically, arithmetic round-off is analyzed by means of a linear-noise model obtained by inserting an additive noise source at each point in the computation algorithm where round-off occurs. However, the effects of round-off noise are very similar among the different classes of FFT algorithms (Oppenheim et al., 1999).

Generally, a FFT processor which computes $N$ samples, represented as $n_{FFT}+1$ bit signed fractions, is affected by a round-off noise which adds to the inaccuracy for transfer impedance, in amplitude (Oppenheim et al., 1999)

$$\left.\frac{\Delta|Z|}{|Z|}\right|_{Round-off} = \frac{N}{2^{n_{FFT}-1}}, \qquad (5.1)$$

and in phase (Dishan, 1995)(Ming and Kang, 1996),

$$\left.\frac{\Delta\Phi_Z}{\Phi_Z}\right|_{Round-off} = \frac{1}{\pi\sqrt{N}\,2^{n_{FFT}}}. \tag{5.2}$$

So, maximizing the register length to $n_{FFT}=32$, the round-off noise is entirely negligible. Once that $n_{FFT}<32$, if the number of samples is increased $N>>1$, then the round-off noise due to FFT degrades the accuracy of transfer impedance, so much more in amplitude (5.1) how much less in phase (5.2). One can overcome this inconvenience by iterating the FFT processor for $A$ cycles, as the best estimate of one FFT value is the average of $A$ FFT repeated values. The improvement is that the inaccuracy for the averaged transfer impedance, in amplitude and phase, consists on the error of quantization due to the uniform sampling ADC (3.4)-(3.7) and the round-off noise due to FFT (5.1)-(5.2), the last term being decreased of $\sqrt{A}$, i.e.:

$$\frac{\Delta|Z|}{|Z|} = \left.\frac{\Delta|Z|}{|Z|}\right|_U + \frac{1}{\sqrt{A}}\left.\frac{\Delta|Z|}{|Z|}\right|_{Round-off} = \frac{1}{2^n} + \frac{1}{\sqrt{A}}\frac{N}{2^{n_{FFT}-1}}, \tag{5.3}$$

$$\frac{\Delta\Phi_Z}{\Phi_Z} = \left.\frac{\Delta\Phi_Z}{\Phi_Z}\right|_U + \frac{1}{\sqrt{A}}\left.\frac{\Delta\Phi_Z}{\Phi_Z}\right|_{Round-off} = 2\frac{f}{f_S} + \frac{1}{\sqrt{A}}\frac{1}{\pi\sqrt{N}\,2^{n_{FFT}}}. \tag{5.4}$$

Once reduced the register length to $n_{FFT}\leq 16$, only if the FFT processor performs the averages during a number of cycles

$$\left(\frac{N}{2^{n_{FFT}-n-1}}\right)^2 << A \leq N^2, \tag{5.5}$$

then the round-off noise due to FFT can be neglected with respect to the quantization error due to uniform ADC, in amplitude

$$\frac{\Delta|Z|}{|Z|} \approx \frac{1}{2^n} + \frac{1}{2^{n_{FFT}-1}} \simeq \left.\frac{\Delta|Z|}{|Z|}\right|_U = \frac{1}{2^n}, \tag{5.6}$$

and especially in phase

$$\frac{\Delta\Phi_Z}{\Phi_Z} \simeq 2\frac{f}{f_S} + \frac{1}{\pi N 2^{n_{FFT}}} \cong \left.\frac{\Delta\Phi_Z}{\Phi_Z}\right|_U = 2\frac{f}{f_S}. \tag{5.7}$$

So, the round-off noise due to FFT is compensated. The quantization error due to ADC decides the accuracy for transfer impedance: it is constant in amplitude, once fixed the bit resolution $n$, and can be limited in phase, by an oversampling technique $f_S >> f$.

In the limit of the Shannon-Nyquist theorem, an electric signal with band of frequency $B$ must be sampled at the minimal rate $f_S = 2B$, holding $N_{min}$ samples in a window of time $T$. Instead, in the hypothesis of oversampling, the signal can be sampled holding the same number of samples $N_{min}$ but in a shorter time window $T/R_O$, due to an high ratio of sampling:

$$R_O = \frac{f_S}{2B} >> 1 . \tag{5.8}$$

This is equivalent to the operating condition in which, during the time window,

$$T = \frac{N_{\min}}{2B} , \tag{5.9}$$

the uniform over-sampling ADC hold a samples number

$$N = R_O \cdot N_{\min} >> N_{\min} \tag{5.10}$$

which is linked to the number of cycles iterated by the FFT processor:

$$A \cong N^2 = R_O^2 \cdot N_{\min}^2 >> 1 . \tag{5.11}$$

As comments on eqs. (5.9)-(5.11), a low number of samples $N_{min}$, corresponding to the Shannon-Nyquist limit, shortens the time window (5.9). An high oversampling ratio lowers the phase inaccuracy although it raises the samples number hold by uniform ADC and especially the cycles number iterated by FFT; however, even a minimal oversampling ratio $R_{O,min}$ limits the phase inaccuracy with the advantage of not raising to much the samples number hold by ADC (5.10) and especially the cycles number iterated by FFT (5.11).

The quadrupole (frequency band *B)* exhibits a galvanic contact with the subjacent non-saturated medium (terrestrial soil or concrete with low permittivity $\varepsilon_r = 4$ and high resistivity $\sigma_S \approx 3.334 \cdot 10^{-4}$ $S/m$, $\sigma_C \approx 10^{-4}$ $S/m$). It is required that the inaccuracy $\Delta\varepsilon_r/\varepsilon_r(f,f_S,n)$ in the measurement of $\varepsilon_r$ is below a prefixed limit $\Delta\varepsilon_r/\varepsilon_r|_{fixed}$ (*10-15%*) within the band *B (100kHz)*. As first result, if the samples number satisfying the Shannon-Nyquist theorem is minimized, i.e. $N_{min}=2$, then the time window for sampling is shortened to $T = N_{min}/(2B) = 1/B \approx 10\mu s$. In order to analyze the complex impedance measured by the probe in Fourier domain, an uniform ADC can be connected to a FFT processor, being affected by a round-off amplitude noise. As conclusive result, a technique of compensation must occur. The ADC must be specified by: a minimal bit resolution $n \leq 12$, thereby rendering the system of measurement almost insensitive to the electric noise of the external environment; and a minimal over-sampling rate $f_S$, which limits the ratio $R_O=f_S/(2B)$, so the actual samples number $N = R_O \cdot N_{min}$ is up to one hundred (soil, $f_S = 10MHz$, $R_O = 50$, $N \approx 100$)(concrete, $f_S = 5MHz$, $R_O = 25$, $N \approx 50$). Moreover, even if the FFT register length is equal to $n_{FFT} = 16$, anyway the minimal rate $f_S$ ensures a number of averaged FFT values $A \leq N^2$ even up to ten thousand, necessary to bound the round-off noise (soil, $A \approx 10^4$)(concrete, $A \approx 2.5 \cdot 10^3$) [fig. 11][tab. 3].

## 6. Characteristic geometrical dimensions of the quadrupolar probe.

In this section, we refer to Vannaroni's paper (Vannaroni et al., 2004) which discusses the dependence of TX current and RX voltage on the array and electrode dimensions. The dimensions of the quadrupolar probe terminals are not critical in the definition of the system because they can be considered point electrodes with respect to their separating distances. In this respect, the separating distance to consider are either the square array side [fig. 12.a] or the spacing distance for a Wenner configuration [fig. 12.b]. The only aspect that could be of importance for the practical implementation of the system is the relationship between electrode dimensions and the magnitude of the current injected into the ground. Current is a critical parameter of the mutual impedance probe in that, in general, given the practical voltage levels applicable to the electrodes and the capacitive coupling with the soil, the current levels are expected to be quite low, with a resulting limit to the accuracy that can be achieved for the amplitude and phase measurements. Furthermore, low currents imply a reduction of the voltage signal read across the RX terminals and more stringent requirements for the reading amplifier.

In the Appendix it is proven that, having fixed the input resistance $R_{in}$ of the amplifier stage and selecting the minimum value of frequency $f_{min}$ for the quadrupole [fig. 1], which allows an inaccuracy in the measurement of the dielectric permittivity $\varepsilon_r$ below a prefixed limit (*10%*), then the radius $r(R_{in},f_{min})$ of the electrodes can be designed, as it depends only on the resistance $R_{in}$ and the frequency $f_{min}$, the radius being an inversely proportional function to both $R_{in}$ and $f_{min}$ [fig. 13][tab. 4],

$$r \simeq \frac{1}{(2\pi)^2 \varepsilon_0 R_{in} f_{\min}}. \tag{6.1}$$

Moreover, known the minimum bit resolution $n_{min}$ for the uniform or IQ sampling ADC, which allows an inaccuracy for permittivity $\varepsilon_r$ below the limit *10%*, the electrode-electrode distance $L(r,n_{min})$ can also be defined, as it depends only on the radius $r(R_{in},f_{min})$ and the bit resolution $n_{min}$, the distance being directly proportional to $r(R_{in},f_{min})$ and increasing as the exponential function $2^{n_{min}}$ of $n_{min}$ [fig. 13][tab. 4],

$$L \propto r \cdot 2^{n_{\min}}. \tag{6.2}$$

Finally, the radius $r(R_{in},f_{min})$ remains invariant whether the probe assumes the linear (Wenner's) or the square configuration, while, having also given the resolution $n_{min}$, then the distance $L_S(r,n_{min})$ in the square configuration must be smaller by a factor ($2-2^{1/2}$) compared to the corresponding distance $L_W(r,n_{min})$ in the Wenner configuration (Settimi et al., 2010)

$$L_S = (2-\sqrt{2}) \cdot L_W = (2-\sqrt{2}) \cdot r \cdot 2^{n_{\min}}. \tag{6.3}$$

## 7. Conclusive discussion.

This paper has discussed the development and engineering of electrical spectroscopy for simultaneous and non invasive measurement of electrical resistivity and dielectric permittivity. A suitable quadrupolar probe is able to perform measurements on a subsurface with inaccuracies below a fixed limit (*10%*) in a bandwidth of low (LF) frequency (*100kHz*). The quadrupole probe should be connected to an appropriate analogical digital converter (ADC) which samples in phase and quadrature (IQ) or in uniform mode. If the quadrupole is characterized by a galvanic contact with the surface, the inaccuracies in the measurement of resistivity and permittivity, due to the IQ or uniform sampling ADC, have been analytically expressed. A large number of numerical simulations has proven that the performances of the probe depend on the selected sampler and that the IQ is better compared to the uniform mode under the same operating conditions, i.e. bit resolution and medium [fig. 6][tab. 1].

As regards the IQ mode, it is specified by an inaccuracy $\Delta\Phi_Z/\Phi_Z(n_{IQ})$ for the phase of the transfer impedance which depends only on the bit resolution $n_{IQ}$, assuming small values [for example $\Delta\Phi_Z/\Phi_Z|_{max}(12)=1/2^{12}\approx 2.4\cdot 10^{-4}$] over the entire frequency band $B=100kHz$ of the quadrupole; in the uniform mode, the corresponding inaccuracy $\Delta\Phi_Z/\Phi_Z(f,f_S)$ depends on the frequency $f$, only being small for frequencies much lower than the sampling rate $f_S$ [i.e. $2f/f_S\leq \Delta\Phi_Z/\Phi_Z|_{max}(12)$]. The principal advantages are: firstly, the minimum value $f_{min}(n)$ of the frequency, which allows an inaccuracy for permittivity $\varepsilon_r$ below *10%*, is slightly lower when the probe is connected to an IQ rather than a uniform ADC, other operating conditions being equal, i.e. the resolution and the surface; and, secondly, the inaccuracy $\Delta\sigma/\sigma$ for conductivity $\sigma$, calculated in $f_{min}(n)$, is smaller using IQ than with uniform sampling, being $\Delta\sigma/\sigma|_{IQ}<\Delta\sigma/\sigma|_U$ of almost one order of magnitude, under the same operating conditions, in particular the resolution. As a minor disadvantage, the optimal frequency $f_{opt}$, which minimizes the inaccuracy for $\varepsilon_r$, is generally higher using IQ than uniform sampling, being $f_{opt,IQ}> f_{opt,U}$ of almost one middle frequency (MF) decade, at most, under the same operating conditions, in particular the surface [fig. 6].

Instead the uniform mode is specified by two degrees of freedom, the resolution of bit $n_U$ and the rate of sampling $f_S$, compared to the IQ mode, characterized by one degree of freedom, the bit resolution $n_{IQ}$. Consequently the quadrupolar probe could be connected to a uniform ADC with a sampling rate $f_S$ sufficiently fast to reach, at a resolution (for example $n_U=8$) lower than the IQ's (i.e. $n_{IQ}=12$), the same prefixed limit (i.e. *10%*) of inaccuracy in the measurement of the dielectric permittivity $\varepsilon_r$, for various media, especially those with low electrical conductivities $\sigma$ [tab.1].

Assuming that the electric current injected in materials and so the voltage measured by probe are quasi-monochromatic signals, i.e. with a very narrow frequency band, an IQ down-sampling process can be employed. Besides the quantization error of IQ ADC, which can be assumed small both in amplitude and phase, as decreasing exponentially with the bit resolution, the electric signals are affected by two additional noises. The amplitude term noise, due to external environment, is modeled by the signal to noise ratio which can be reduced performing averages over a thousand of repeated measurements. The phase term noise, due to a phase-splitter detector, which, even if increasing linearly with the frequency, can be minimized by digital electronics providing a rise time of few nanoseconds [figs. 9,10] [tab. 2].

Instead, in order to analyze the complex impedance measured by the quadrupole in Fourier domain, an uniform ADC, which is characterized by a sensible phase inaccuracy depending on frequency, must be connected to a Fast Fourier Transform (FFT) processor, that is especially affected by a round-off amplitude noise linked to both the FFT register length and samples number. If the register length is equal to *32* bits, then the round-off noise is entirely negligible, else, once bits are reduced to *16*, a technique of compensation must occur. In fact, oversampling can be employed within a short time window, reaching a compromise between the needs of limiting the phase inaccuracy due to ADC and not raising too much the number of averaged FFT values sufficient to bound the round-off [fig. 11][tab. 3].

Since, in this paper, the conceptual and technical problem for designing the "heart" of the instrument have already been addressed and resolved, a further paper will complete the technical project, focusing on the following two aims.

First aim: the implementation of hardware which can handle numerous electrodes, arranged so as to provide data which is related to various depths of investigation for a single measurement pass;

consequently, this hardware must be able to automatically switch transmitting and measurement pairs.

Second aim: the implementation of acquisition configurations, by an appropriate choice of transmission frequency, for the different applications in which this instrument can be profitably used.

## Appendix.

A series of two spherical capacitors with radius $r$ and spacing distance $L>>r$ is characterized by the electrical capacitance

$$C = \frac{2\pi\varepsilon_0}{\frac{1}{r} - \frac{1}{L-r}} \simeq 2\pi\varepsilon_0 r \quad , \quad for \quad L >> r \, , \qquad (A.1)$$

with $\varepsilon_0$ the dielectric constant in vacuum.

A quadrupolar probe, with four spherical electrodes of radius $r$ and separating distance $L>>r$, is arranged in the Wenner's configuration, with total length $L_{tot}=3L$, which is specified by the pairs of transmitting electrodes $T_1$ and $T_2$ at the ends of quadrupole and the reading electrodes $R_1$ and $R_2$ in the middle of probe. The quadrupolar probe is characterized by a capacitance almost invariant for the pairs of electrodes $T_{1,2}$ and $R_{1,2}$,

$$C_{T_1,T_2} = \frac{2\pi\varepsilon_0}{\frac{1}{r} - \frac{1}{3L-r}} \simeq C_{R_1,R_2} = \frac{2\pi\varepsilon_0}{\frac{1}{r} - \frac{1}{L-r}} \simeq C = 2\pi\varepsilon_0 r \quad , \quad for \quad L >> r \, . \qquad (A.2)$$

The charge $Q$ of the electrodes being equal, the electrical voltage across the pair $T_1$ and $T_2$ approximates the voltage between $R_1$ and $R_2$,

$$\Delta V_{T_1,T_2} \simeq \Delta V_{R_1,R_2} \simeq \frac{Q}{C} \, . \qquad (A.3)$$

As regards the equivalent capacitance circuit which schematizes the transmission stage of quadrupole [fig. 14.a], if the effect of the capacitance $C$, across the electrodes, is predominant relative to the shunted capacitances $C_{T1}$ and $C_{T2}$, describing the electrical coupling between the transmitting electrodes and the subjacent medium,

$$C << C_{T_1} = C_{T_2}\,, \tag{A.4}$$

then, the probe, working in the frequency *f*, injects in the medium a minimum bound for the modulus of the current $|I|_{min}$,

$$\left|I\right|_{\min} \simeq \omega\cdot C\cdot\Delta V_{T_1T_2} = 2\pi f\cdot 2\pi\varepsilon_0 r\cdot\Delta V_{T_1,T_2} = (2\pi)^2\varepsilon_0 r f\Delta V_{T_1,T_2}\,, \tag{A.5}$$

$|I|_{min}$ increasing linearly with *f*.

Concerning the equivalent capacitance circuit which represents the reception stage of the quadrupolar probe [fig. 14.b], the effect of the capacitance *C*, across the electrodes, is predominant even relative to the shunted capacitances $C_{R1}$ and $C_{R2}$, describing the coupling between the reading electrodes and the subjacent medium,

$$C << C_{R_1} = C_{R_2}\,. \tag{A.6}$$

If the quadrupole, of electrode-electrode distance *L*, is immersed in vacuum, then it can be characterized by the vacuum capacitance

$$C_0 = 4\pi\varepsilon_0 L\,, \tag{A.7}$$

and measures, in the frequency *f*, a minimum limit $|Z|_{min}$ for the transfer impedance in modulus,

$$\left|Z\right|_{\min} = \frac{1}{\omega\cdot C_0} = \frac{1}{2\pi f\cdot 4\pi\varepsilon_0 L} = \frac{1}{2(2\pi)^2\varepsilon_0 L f}\,, \tag{A.8}$$

which gives rises to a minimum for the electrical voltage $\Delta V_{min,R1,R2}$, flowing the current $|I|_{min}$,

$$\Delta V^{\min}_{R_1,R_2} = \left|Z\right|_{\min}\left|I\right|_{\min} \simeq \frac{1}{2(2\pi)^2\varepsilon_0 L f}\cdot(2\pi)^2\varepsilon_0 r f\Delta V_{T_1,T_2} = \frac{1}{2}\frac{r}{L}\Delta V_{T_1,T_2}\,. \tag{A.9}$$

Notice that $\Delta V_{min,R1,R2}$ is independent of *f*, as $|I|_{min}$ is directly and $|Z|_{min}$ inversely proportional to *f*. As a first finding, the analogical digital converter (ADC), downstream of probe, must be specified by a bit resolution *n*, such that:

$$\frac{\Delta V^{\min}_{R_1,R_2}}{\Delta V_{T_1,T_2}} \simeq \frac{1}{2}\frac{r}{L} \approx \frac{\Delta V^{\min}_{R_1,R_2}}{\Delta V_{R_1,R_2}} = \frac{1}{2^{n+1}} \quad\Rightarrow\quad \frac{r}{L} \approx \frac{1}{2^n}\,. \tag{A.10}$$

Instead, if the quadrupole exhibits a galvanic contact with a medium of electrical conductivity $\sigma$ and dielectric permittivity $\varepsilon_r$, working in a band lower than the cut-off frequency, i.e. $f_T = f_T(\sigma,\varepsilon_r) = 1/2\pi\cdot\sigma/\varepsilon_0(\varepsilon_r+1)$, then it measures the transfer impedance in modulus

$$|Z| = \frac{1}{2\pi\sigma L}. \tag{A.11}$$

As final findings, the voltage amplifier, downstream of the probe, must be specified by an input resistance $R_{in}$ larger than both the transfer impedance, i.e.

$$R_{in} > |Z| = \frac{1}{2\pi\sigma L} \quad \Rightarrow \quad L > \frac{1}{2\pi\sigma R_{in}}, \tag{A.12}$$

and the reactance associated to the capacitance $C$, which is characterized by a maximum value in the minimum of frequency $f_{min}$, i.e.

$$R_{in} > \frac{1}{\omega_{\min} C} = \frac{1}{2\pi f_{\min} \cdot 2\pi\varepsilon_0 r} = \frac{1}{(2\pi)^2 \varepsilon_0 r f_{\min}} \quad \Rightarrow \quad r > \frac{1}{(2\pi)^2 \varepsilon_0 R_{in} f_{\min}}. \tag{A.13}$$

## References.

Al-Qadi I. L., O. A. Hazim, W. Su and S. M. Riad (1995). Dielectric properties of Portland cement concrete at low radio frequencies, J. Mater. Civil. Eng., 7, 192-198.

Andren C. and J. Fakatselis (1995). Digital IF Sub Sampling Using the HI5702, HSP45116 and HSP43220, App. Note (Harris DSP and Data Acq.), No. AN9509.1 .

Arpaia P., P. Daponte and L. Michaeli (1999). Influence of the architecture on ADC error modelling, IEEE T. Instrum. Meas, 48, 956-966.

Arpaia P., P. Daponte and S.Rapuano (2003). A state of the art on ADC modelling, Comput. Stand. Int., 26, 31–42.

Auty R.P. and R.H.Cole (1952). Dielectric properties of ice and solid, J. Chem. Phys., 20, 1309-1314.

Björsell N. and P. Händel (2008). Achievable ADC performance by post-correction utilizing dynamic modeling of the integral nonlinearity, Eurasip J. Adv. Sig. Pr., 2008, ID 497187 (10 pp).

Chelidze T.L. and Y. Gueguen (1999). Electrical spectroscopy of porous rocks: a review-I, Theoretical models, Geophys. J. Int., 137, 1-15.

Chelidze T.L. and Y. Gueguen, C. Ruffet (1999). Electrical spectroscopy of porous rocks: a review-II, Experimental results and interpretation, Geophys. J. Int., 137, 16-34.

Debye P. (1929). Polar Molecules (Leipzig Press, Germany).

Declerk P. (1995). Bibliographic study of georadar principles, applications, advantages, and inconvenience, NDT & E Int., 28, 390-442 (in French, English abstract).

Del Vento D. and G. Vannaroni (2005). Evaluation of a mutual impedance probe to search for water ice in the Martian shallow subsoil, Rev. Sci. Instrum., 76, 084504 (1-8).

Dishan H. (1995). Phase Error in Fast Fourier Transform Analysis, Mech. Syst. Signal Pr., 9, 113-118.

Edwards R. J. (1998). Typical Soil Characteristics of Various Terrains, http://www.smeter.net/grounds/soil-electrical-resistance.php.

Grard R. (1990). A quadrupolar array for measuring the complex permittivity of the ground: application to earth prospection and planetary exploration, Meas. Sci. Technol., 1, 295-301.

Grard R. (1990). A quadrupole system for measuring in situ the complex permittvity of materials: application to penetrators and landers for planetary exploration, Meas. Sci. Technol., 1, 801-806.

Grard R. and A. Tabbagh (1991). A mobile four electrode array and its application to the electrical survey of planetary grounds at shallow depth, J. Geophys. Res., 96, 4117-4123.

Jankovic D. and J. Öhman (2001). Extraction of in-phase and quadrature components by IF-sampling, Department of Signals and Systems, Cahlmers University of Technology, Goteborg (carried out at Ericson Microwave System AB).

Knight R. J. and A. Nur (1987). The dielectric constant of sandstone, 60 kHz to 4 MHz, Geophysics, 52, 644-654.

Kuffel J., R. Malewsky and R. G. Van Heeswijk (1991). Modelling of the dynamic performance of transient recorders used for high voltage impulse tests, IEEE T. Power Deliver., 6, 507-515.

Laurents S., J. P. Balayssac, J. Rhazi, G. Klysz and G. Arliguie (2005). Non-destructive evaluation of concrete moisture by GPR: experimental study and direct modeling, Mater. Struct., 38, 827-832.

Ming X. and D. Kang (1996). Corrections for frequency, amplitude and phase in Fast Fourier transform of harmonic signal, Mech. Syst. Signal Pr., 10, 211-221.

Mojid M. A., G. C. L. Wyseure and D. A. Rose (2003). Electrical conductivity problems associated with time-domain reflectometry (TDR) measurement in geotechnical engineering, Geotech. Geo. Eng., 21, 243-258.

Mojid M. A. and H. Cho (2004). Evaluation of the time-domain reflectometry (TDR)-measured composite dielectric constant of root-mixed soils for estimating soil-water content and root density, J. Hydrol., 295, 263–275.

Murray-Smith D. J. (1987). Investigations of methods for the direct assessment of parameter sensitivity in linear closed-loop control systems, in Complex and distributed systems: analysis, simulation and control, edited by Tzafestas S. G. and Borne P. (North-Holland, Amsterdam), pp. 323–328.

Myounghak O., K. Yongsung and P. Junboum (2007). Factors affecting the complex permittivity spectrum of soil at a low frequency range of 1 kHz10 MHz, Environ. Geol., 51, 821-833.

Oppenheim A. V., R.W. Schafer and J. R. Buck (1999). Discrete-Time Signal Processing (Prentice Hall International, Inc., New York - II Ed.).

Polder R., C. Andrade, B. Elsener, Ø. Vennesland, J. Gulikers, R. Weidert and M. Raupach (2000). Test methods for on site measurements of resistivity of concretes, Mater. Struct., 33, 603-611.

Polge R. J., B. K. Bhagavan and L. Callas (1975). Evaluating analog-to-digital converters, Simulation, 24, 81-86.

Razavi B. ( 1995). Principles of Data Conversion System Design (IEEE Press, New York).

Samouëlian A., I. Cousin, A. Tabbagh, A. Bruand and G. Richard (2005). Electrical resistivity survey in soil science: a review, Soil Till,. Res., 83, 172-193.

Sbartaï Z. M., S. Laurens, J. P. Balayssac, G. Arliguie and G. Ballivy (2006). Ability of the direct wave of radar ground-coupled antenna for NDT of concrete structures, NDT & E Int., 39, 400-407.

Settimi A., A. Zirizzotti, J. A. Baskaradas and C. Bianchi (April 2010). Inaccuracy assessment for simultaneous measurements of resistivity and permittivity applying sensitivity and transfer function approaches, Ann. Geophys. – Italy, 53, 2, 1-19; *ibid.,* Earth-prints, http://hdl.handle.net/2122/5180 (2009); *ibid.,* arXiv:0908.0641 [physics.geophysiscs] (2009).

Tabbagh A., A. Hesse and R. Grard (1993). Determination of electrical properties of the ground at shallow depth with an electrostatic quadrupole: field trials on archaeological sites, Geophys. Prospect., 41, 579-597.

Vannaroni G., E. Pettinelli, C. Ottonello, A. Cereti, G. Della Monica, D. Del Vento, A. M. Di Lellis, R. Di Maio, R. Filippini, A. Galli, A. Menghini, R. Orosei, S. Orsini, S. Pagnan, F. Paolucci, A. Pisani R., G. Schettini, M. Storini and G. Tacconi (2004). MUSES: multi-sensor soil electromagnetic sounding, Planet. Space Sci., 52, 67–78.

Zhang J. Q. and S. J. Ovaska (1998). ADC characterization by an eigenvalues method, Instrumentation and Measurement Technology Conference (IEEE), 2, 1198-1202.

**Tables and captions.**

Table 1.a

| ***Soil*** ***($\varepsilon_r = 4$, $\rho = 3000\ \Omega\cdot m$)*** | **U Sampling ADC (12 bit, 10 MS/s)** | **U Sampling (12 bit, 60 MS/s)** | **(12 bit, 200 MS/s)** | **IQ Sampling (12 bit)** |
|---|---|---|---|---|
| $f_{opt}$ | 150.361 kHz | 267.569 kHz | 387.772 kHz | ≈ 1.459 MHz |
| $f_{min}$ | 92.198 kHz | 98.042 kHz | 95.055 kHz | 94.228 kHz |
| $f_{max}$ | 265.287 kHz | 885.632 kHz | ≈ 1.832 MHz | ≈ 16.224 MHz |

Table 1.b

| ***Concrete*** ***($\varepsilon_r = 4$, $\rho = 10000\ \Omega\cdot m$)*** | **U Sampling ADC (8 bit, 50 MS/s)** | **U Sampling (8 bit, 100 MS/s)** | **(8 bit, 250 MS/s)** | **(8 bit, 2 GS/s)** |
|---|---|---|---|---|
| $f_{opt}$ | 241.41 kHz | 289.103 kHz | 364.485 kHz | 607.522 kHz |
| $f_{min}$ | 98.574 kHz | 96.629 kHz | 95.553 kHz | 94.953 kHz |
| $f_{max}$ | 591.657 kHz | 815.1 kHz | ≈ 1.079 MHz | ≈ 1.443 MHz |

Table 1.b.bis

| ***Concrete*** ***($\varepsilon_r = 4$, $\rho = 10000\ \Omega\cdot m$)*** | **U Sampling ADC (12 bit, 5 MS/s)** | **U Sampling (12 bit, 10 MS/s)** | **(12 bit, 60 MS/s)** | **(12 bit, 200 MS/s)** | **IQ Sampling (12 bit)** |
|---|---|---|---|---|---|
| $f_{opt}$ | 53.271 kHz | 66.605 kHz | 116.332 kHz | 165.329 kHz | 437.716 kHz |
| $f_{min}$ | 35.085 kHz | 30.646 kHz | 28.517 kHz | 28.273 kHz | 28.268 kHz |
| $f_{max}$ | 85.396 kHz | 169.271 kHz | 549.595 kHz | ≈ 1.016 MHz | ≈ 4.867 MHz |

Table 2

| $f_{opt}$ $f_{S,opt}$<br>$f_{min}$ $f_{S,min}$<br>$f_{max}$ $f_{S,max}$ | **IQ Down - Sampling (n = 12)**<br>**f = M·$f_S$ (M = 8)** | | **+ Electric Noise (SNR = 30 dB, A = $10^3$) + Phase Shifter (τ = 1 ns)** | | **+ Electric Noise (SNR = 30 dB, A = $10^3$) + Phase Shifter ($\Delta\Phi_Z/\Phi_Z$ = 0.2°)** | |
|---|---|---|---|---|---|---|
| *Soil (ε$_r$ = 4, ρ = 3000 Ω·m)* | 1.459 MHz | 182.381 kHz | 775.286 kHz | 96.911 kHz | 733.816 kHz | 91.727 kHz |
| | 94.228 kHz | 11.779 kHz | 105.952 kHz | 13.244 kHz | 110.857 kHz | 13.857 kHz |
| | 16.224 MHz | 2.028 MHz | 5.659 MHz | 707.414 kHz | 7.968 MHz | 995.938 kHz |
| *Concrete (ε$_r$ = 4, ρ = 10000 Ω·m)* | 437.716 MHz | 54.714 kHz | 304.818 kHz | 38.102 kHz | 220.145 kHz | 27.518 kHz |
| | 28.268 kHz | 3.534 kHz | 31.756 kHz | 3.969 kHz | 33.257 kHz | 4.157 kHz |
| | 4.867 MHz | 608.413 kHz | 2.699 MHz | 337.416 kHz | 2.39 MHz | 298.781 kHz |

Table 3.a

| ***Soil***<br>***($\varepsilon_r$ = 4,***<br>***$\rho$ = 3000 Ω·m)*** | **U Sampling ADC**<br>**(n = 12,**<br>**$f_S$ = 10 MHz)**<br>**+**<br>**FFT ($n_{FFT}$ = 16)** |
|---|---|
| $T = N_{min}/2B$ | ≈ 10 µs |
| $N = N_{min}\cdot(f_S/2B)$ | ≈ 100 ($2^7 = 128$) |
| $A \leq N^2$ | ≈ $10^4$ |
| $f_{opt}, f_{min}, f_{max}$<br>$(\Delta\varepsilon_r/\varepsilon_r, \Delta\sigma/\sigma \leq 0.15)$ | 156.256 kHz, 99.68 kHz, 261.559 kHz |

Table 3.b

| ***Concrete***<br>***($\varepsilon_r$ = 4,***<br>***$\rho$ = 10000 Ω·m)*** | **U Sampling ADC**<br>**(n = 8,**<br>**$f_S$ = 50 MHz)**<br>**+**<br>**FFT ($n_{FFT}$ = 16)** | **U Sampling ADC**<br>**(n = 12,**<br>**$f_S$ = 5 MHz)**<br>**+**<br>**FFT ($n_{FFT}$ = 16)** |
|---|---|---|
| $T = N_{min}/2B$ | ≈ 10 µs | |
| $N = N_{min}\cdot(f_S/2B)$ | ≈ 500 ($2^9 = 512$) | ≈ 50 ($2^6 = 64$) |
| $A \leq N^2$ | ≈ $2.5\cdot10^5$ | ≈ $2.5\cdot10^3$ |
| $f_{opt}, f_{min}, f_{max}$ | 241.906 kHz, 99.007kHz, 591.411 kHz<br>$(\Delta\varepsilon_r/\varepsilon_r, \Delta\sigma/\sigma \leq 0.15)$ | 55.344 kHz, 38.195kHz, 83.642 kHz<br>$(\Delta\varepsilon_r/\varepsilon_r, \Delta\sigma/\sigma \leq 0.1)$ |

Table 4.a

| *Soil* *($\varepsilon_r = 4$, $\rho = 3000\ \Omega\cdot m$)* | **Square configuration** | **(Wenner's) Linear configuration** |
|---|---|---|
| $L$, $L_{TOT} = 3\cdot L$ | L ≈ 1 m | $L_{TOT}$ ≈ 10 m |
| $R_{in}$ | 72.8 MΩ | 37.3 MΩ |
| $r$ | 417.042 µm | 813.959 µm |

Table 4.b

| *Concrete* *($\varepsilon_r = 4$, $\rho = 10000\ \Omega\cdot m$)* | **Square configuration** | **(Wenner's) Linear configuration** |
|---|---|---|
| $L$, $L_{TOT} = 3\cdot L$ | L ≈ 1 m | $L_{TOT}$ ≈ 1 m |
| $R_{in}$ | 242 MΩ | 1.24 GΩ |
| $r$ | 418.197 µm | 81.616 µm |

Tab. 1. Refer to the caption of fig. 6: a quadrupolar probe is connected to an uniform or IQ ADC (bit resolution $n$). Optimal working frequency $f_{opt}$, which minimizes the inaccuracy in the measurement of permittivity $\Delta\varepsilon_r/\varepsilon_r(f)$, minimum and maximum frequencies, $f_{min}$ and $f_{max}$, which limit the inaccuracies of permittivity and conductivity, $\Delta\varepsilon_r/\varepsilon_r(f)\leq 0.1$ and $\Delta\sigma/\sigma(f)\leq 0.1$, for measurements performed on: soil, with $n=12$ (a); and concrete, with $n=8$ (b) or $n=12$ (b.bis).

Tab. 2. Refer to the captions of figs. 9 and 10. A quadrupole is connected to a noisy IQ ADC (down-sampling factor $M$). Sampling frequencies, $f_S=f/M$, relative to the optimal, minimum and maximum working frequencies, i.e. $f_{opt}$, $f_{min}$ and $f_{max}$, for measurements performed on soil and concrete.

Tab. 3. Refer to the caption of fig. 11. A probe is connected to an uniform ADC (Shannon-Nyquist theorem: limit of samples number, $N_{min} = 2$), in addition to a FFT processor with round-off noise ($T$, time window; $N$, actual samples number; $A$, cycles number averaging FFT values). Optimal, minimum and maximum working frequencies, $f_{opt}$, $f_{min}$ and $f_{max}$, for measurements performed on soil (a) and concrete (b).

Tab. 4. Radius $r$ of the probe electrodes and characteristic geometrical dimension of the quadrupole in the square (side $L$) or Wenner's (length $L_{TOT}=3L$) configurations, employing an amplifier stage with input resistance $R_{in}$, and that is connected to an IQ sampling ADC with bit resolution $n=12$, in order to perform measurements on soil (a) or concrete (b).

**Figures and captions.**

Figure 1.a.

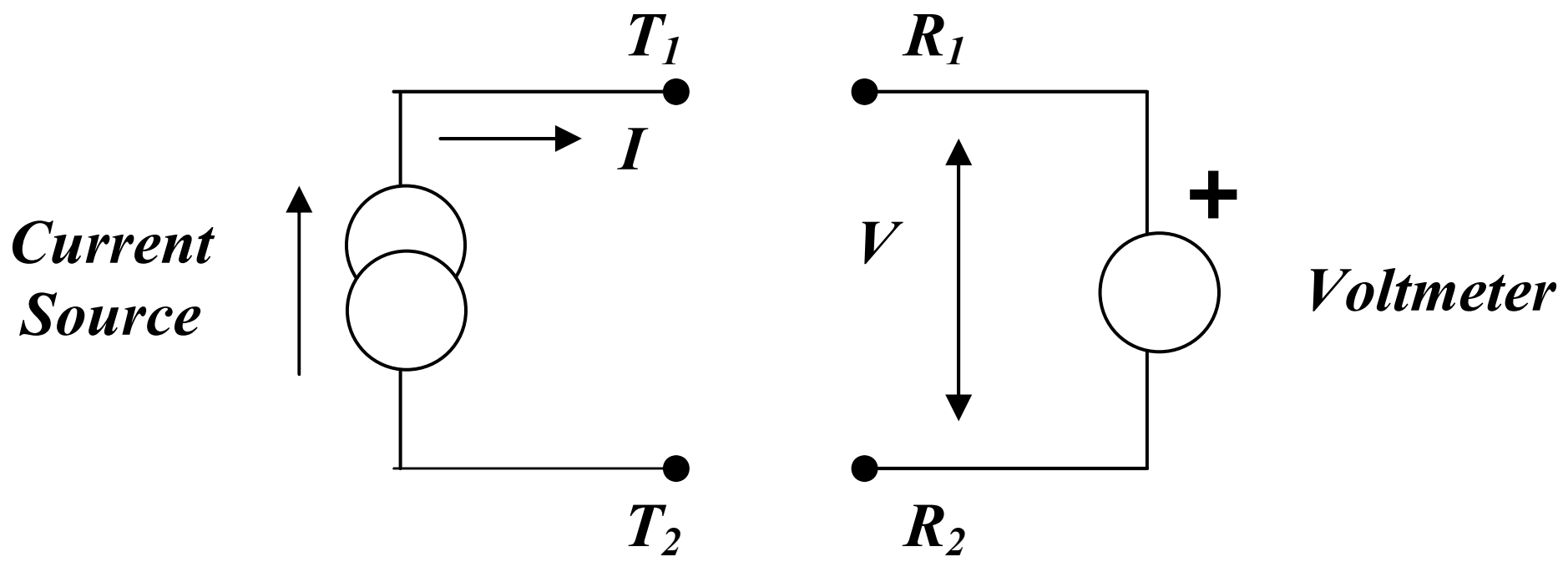

Figure 1.b.

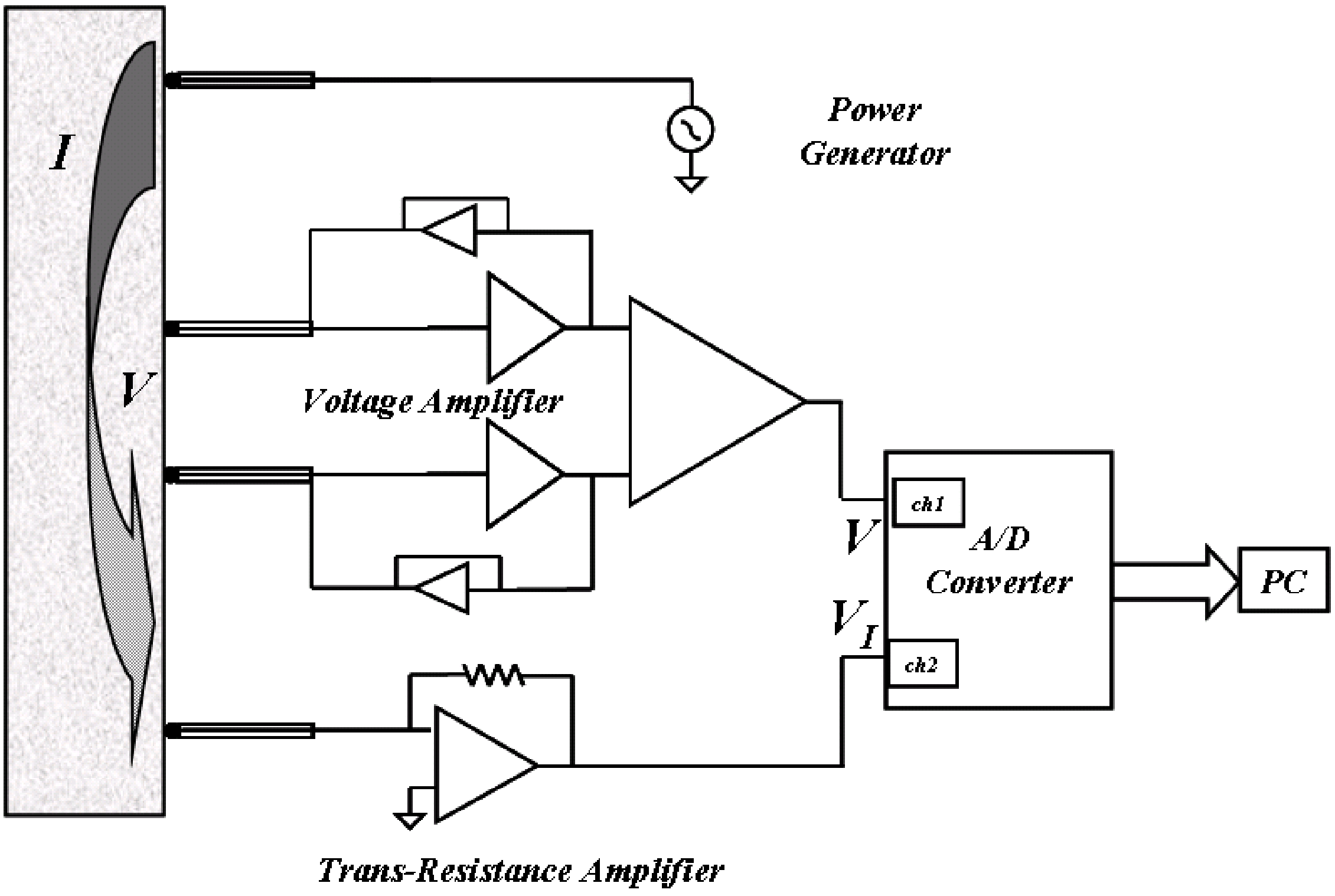

Figure 2.a.

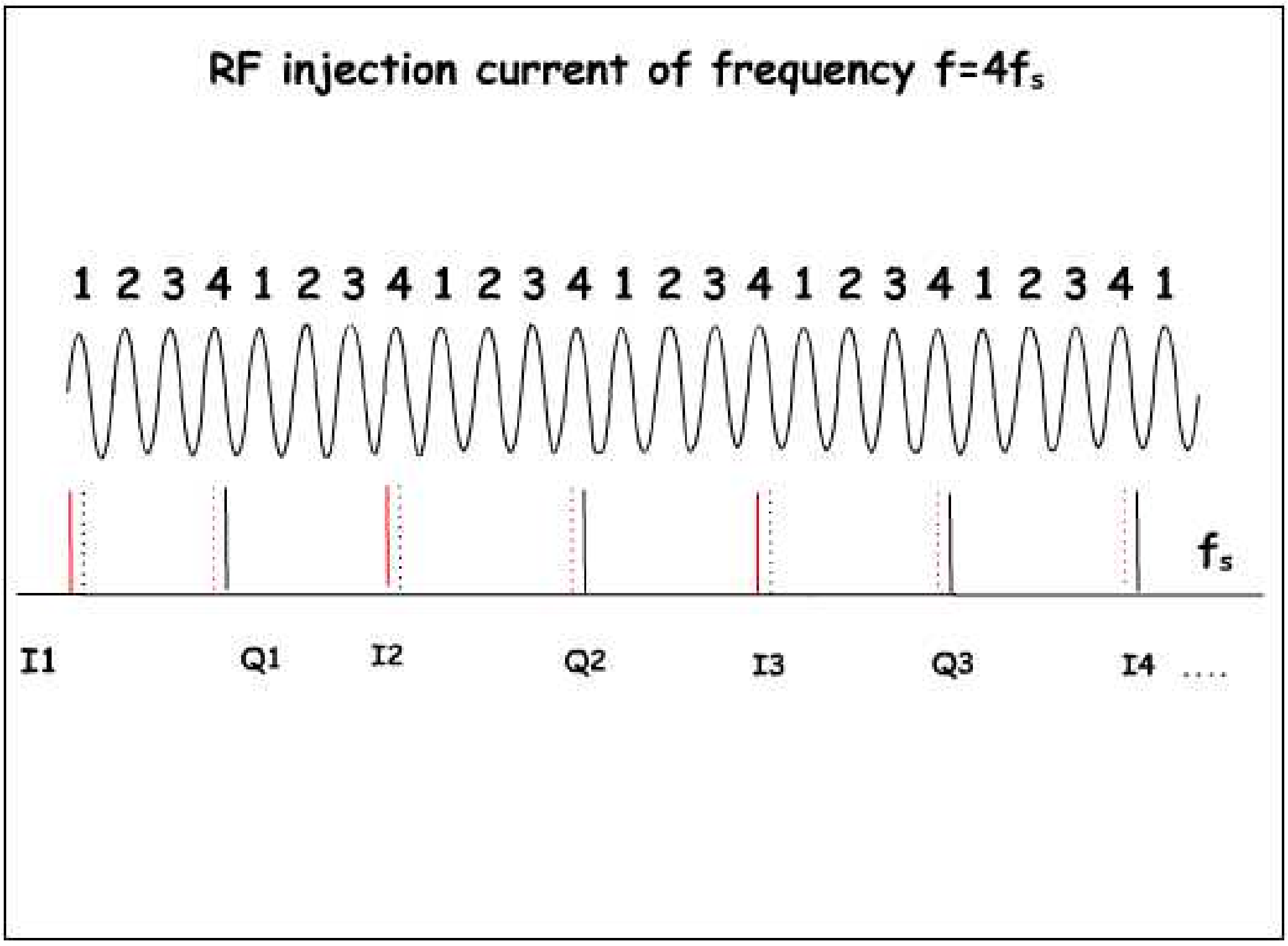

Figure 2.b.

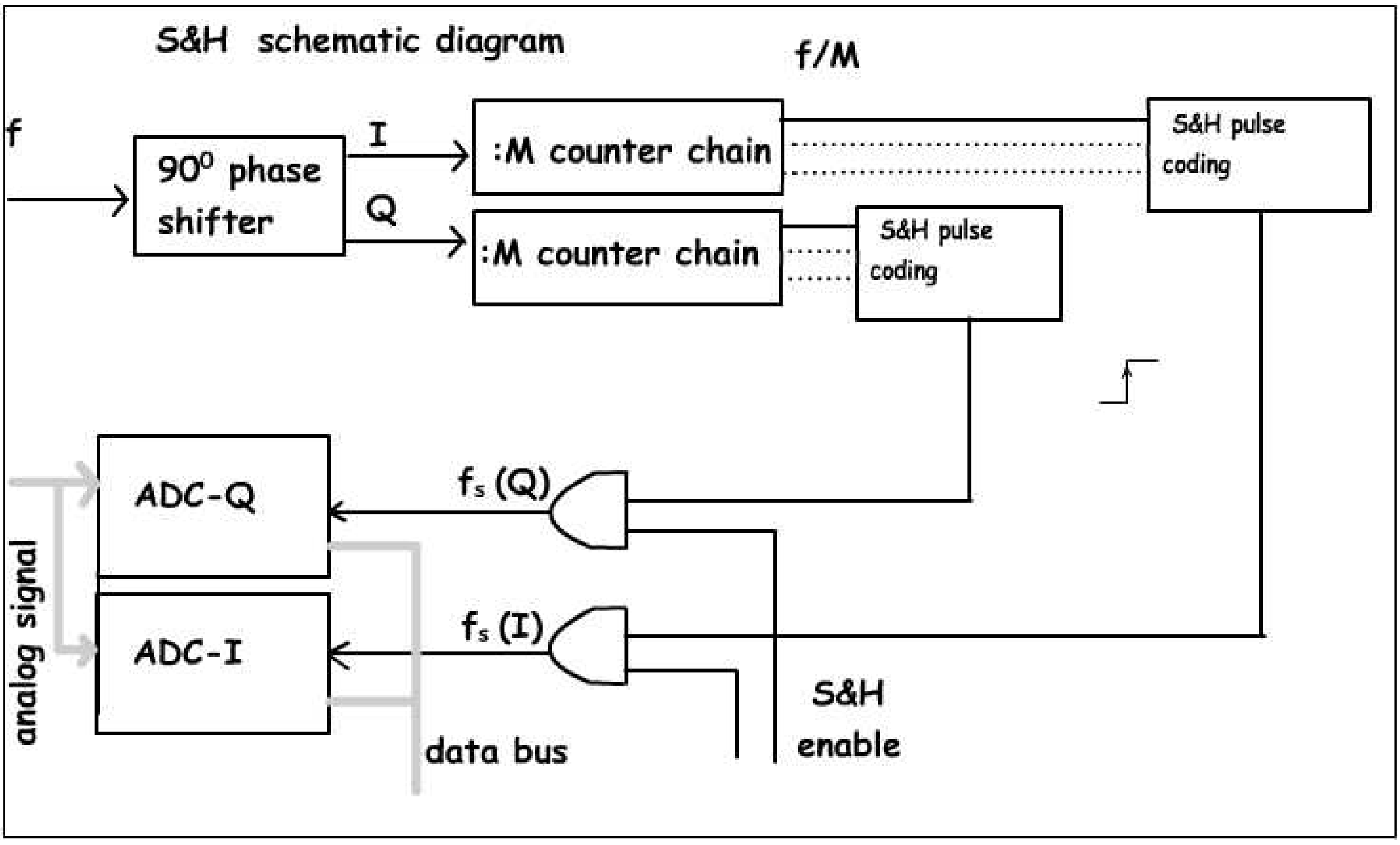

Figure 3.a

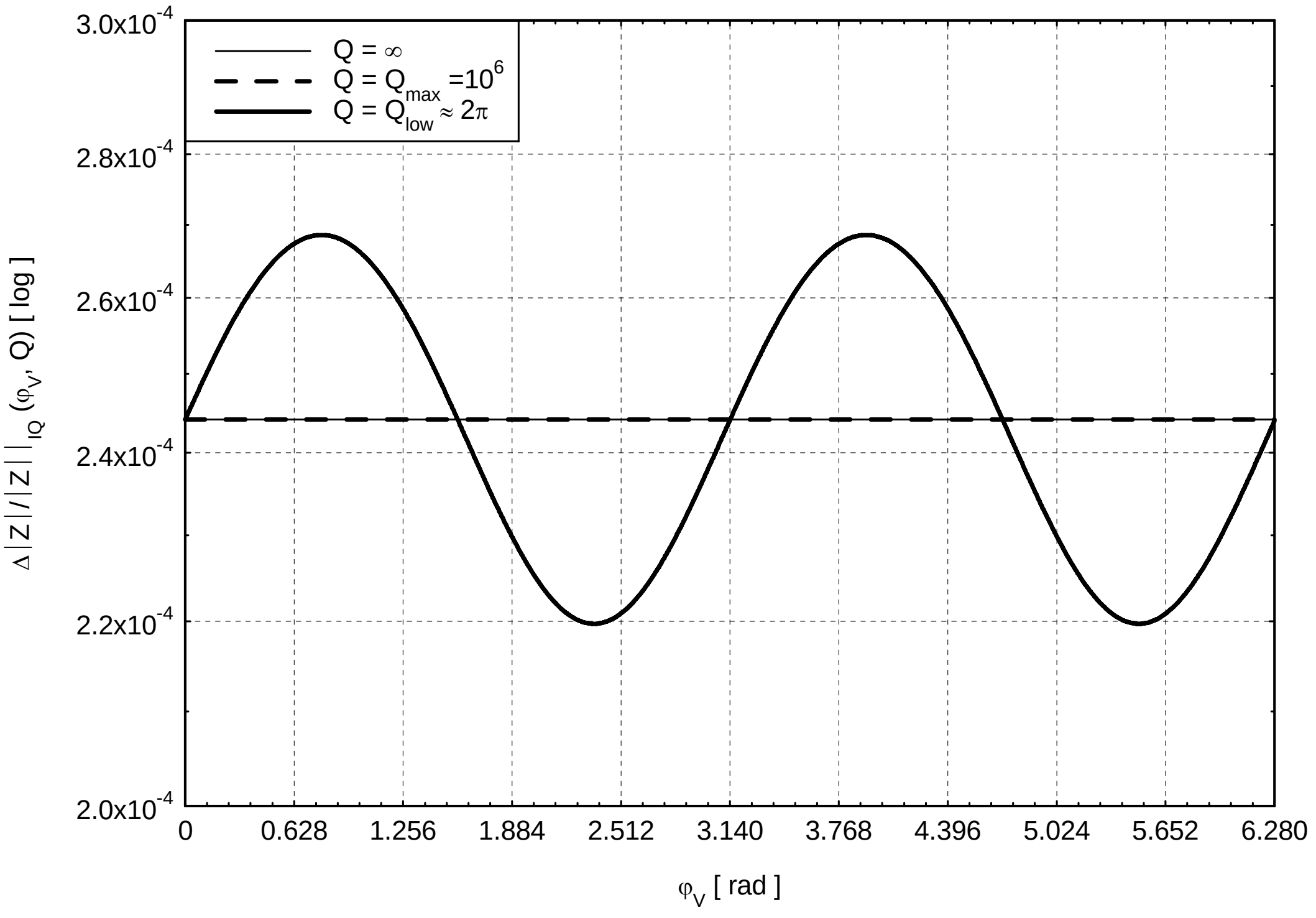

Figure 3.a.bis

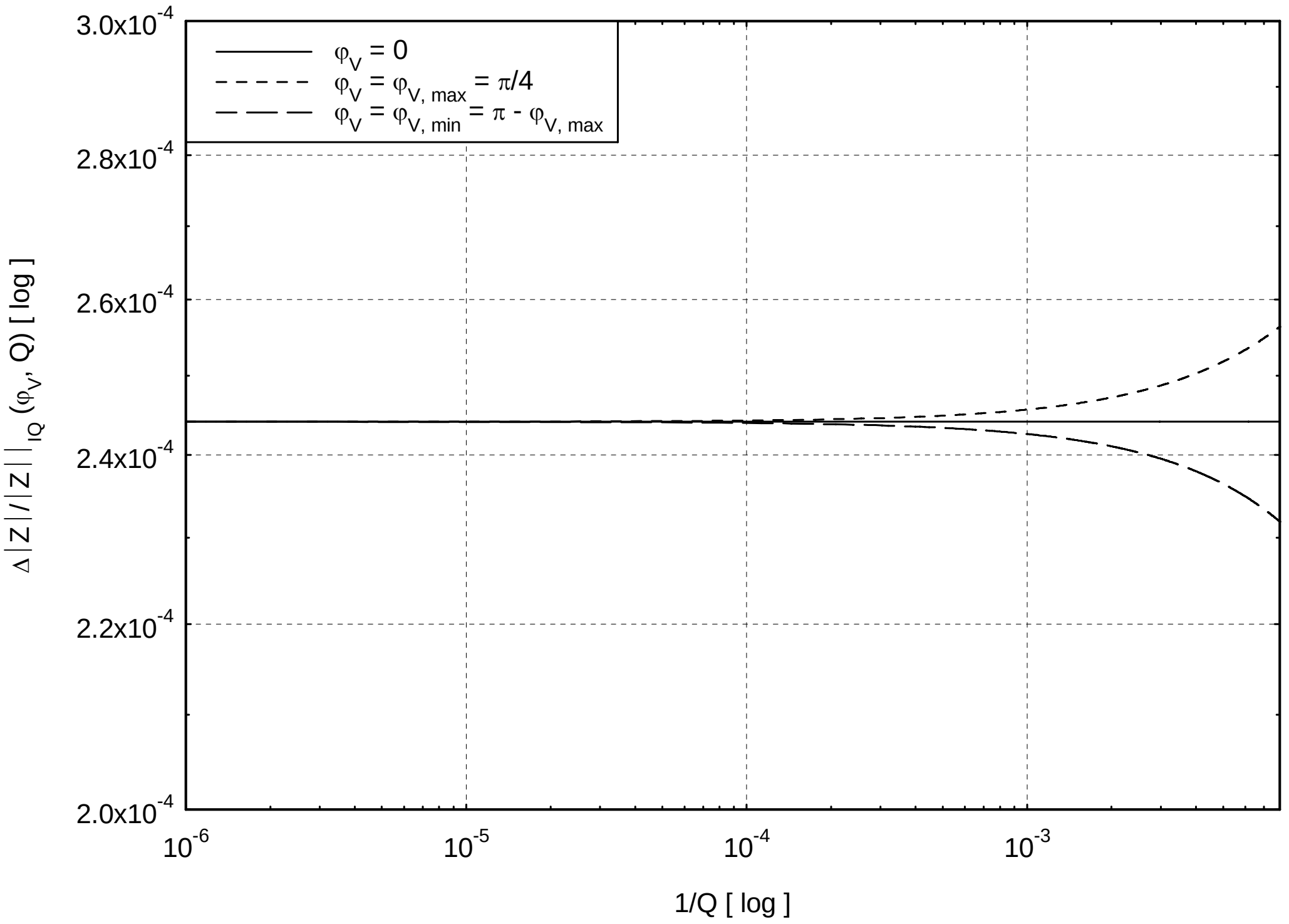

Figure 3.b

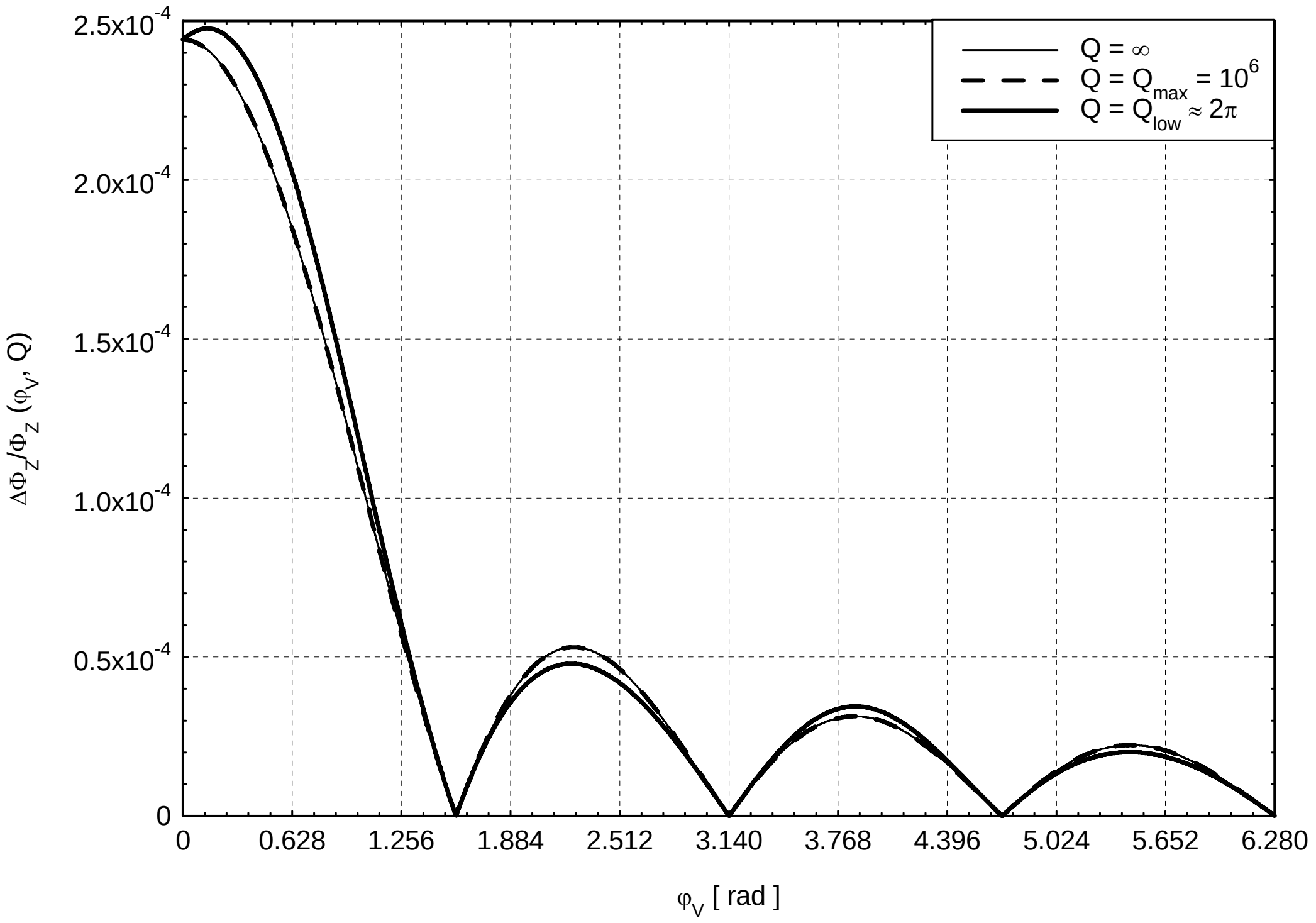

Figure 3.b.bis

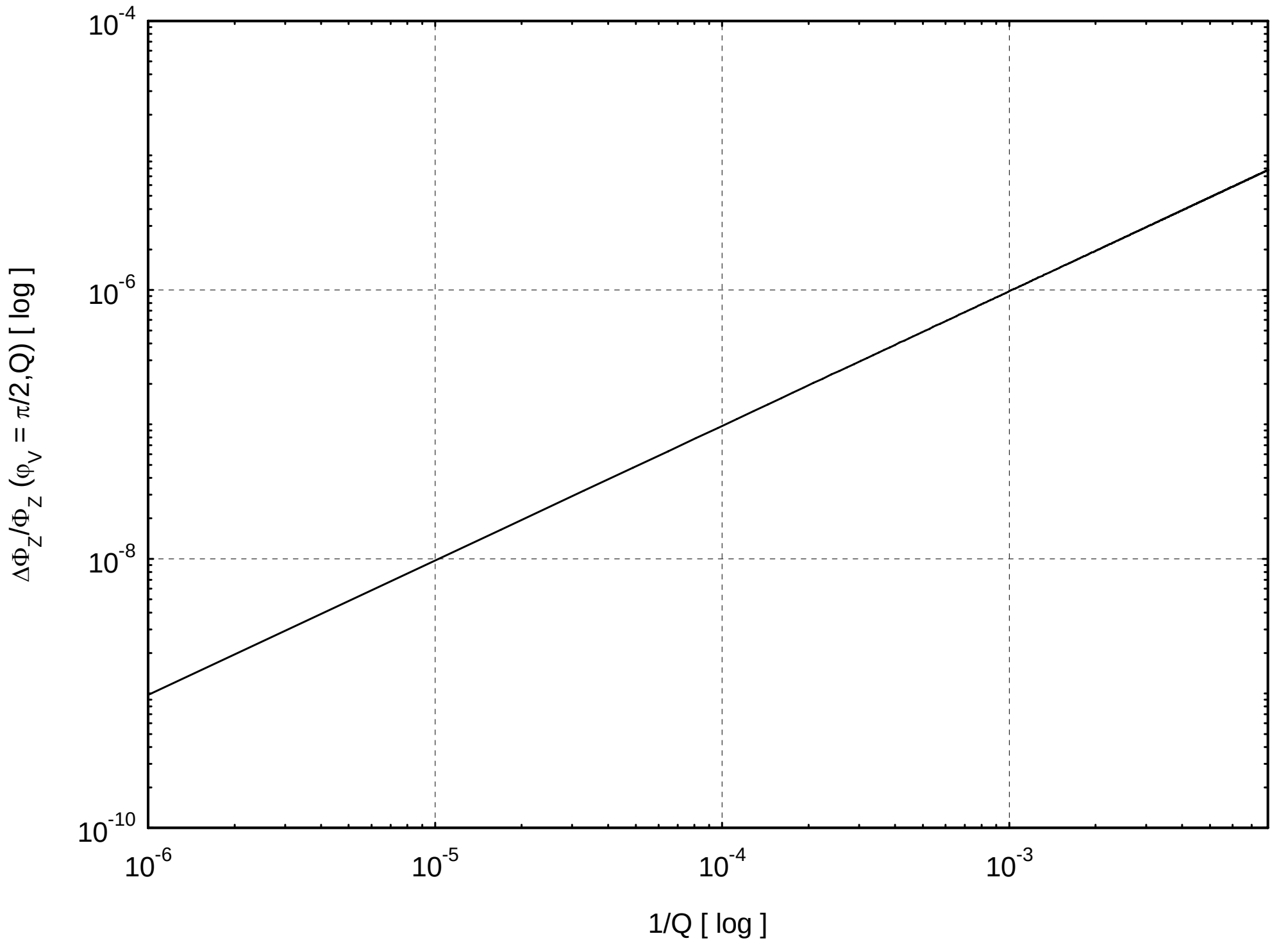

Figure 4.a

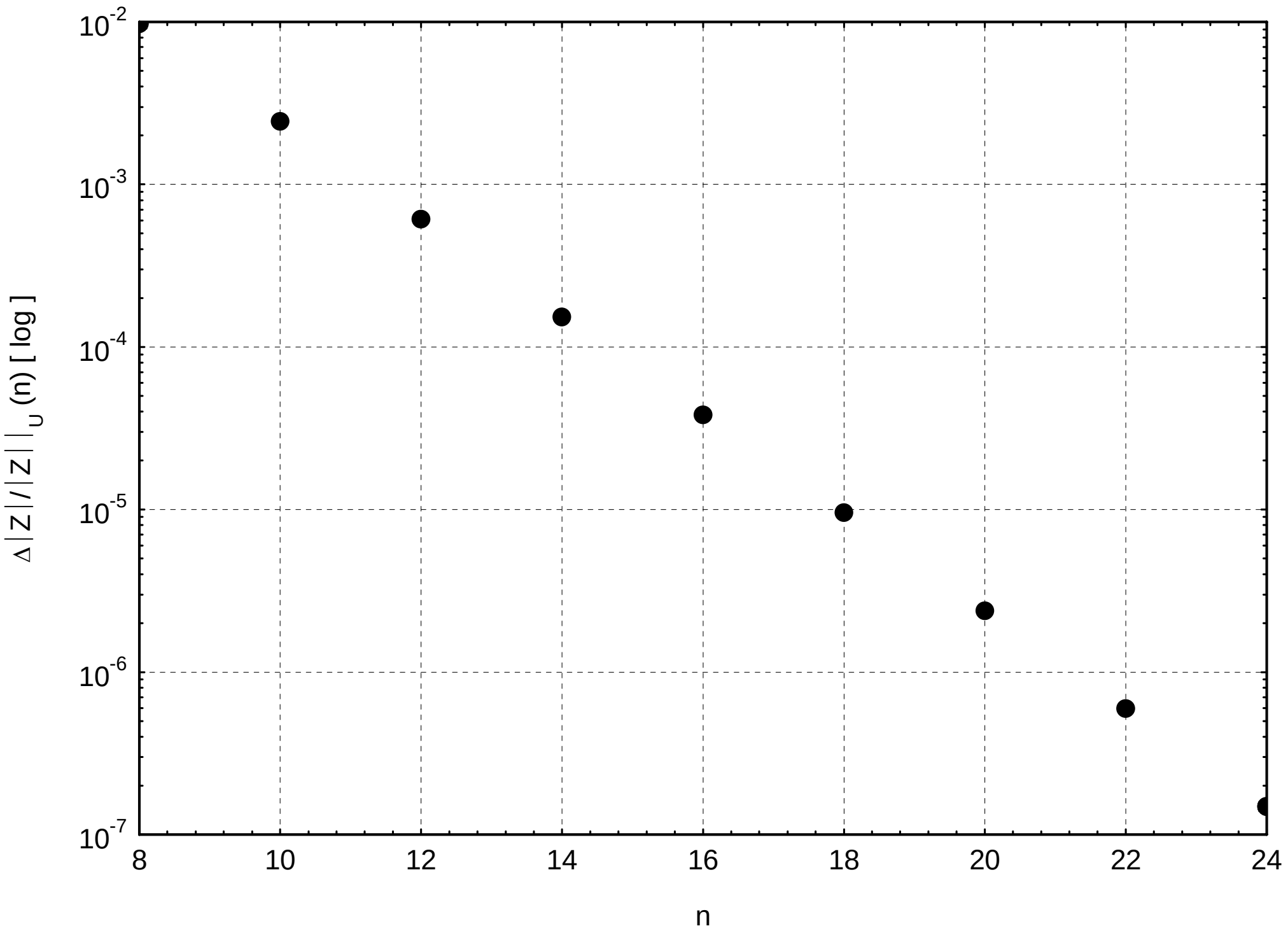

Figure 4.b

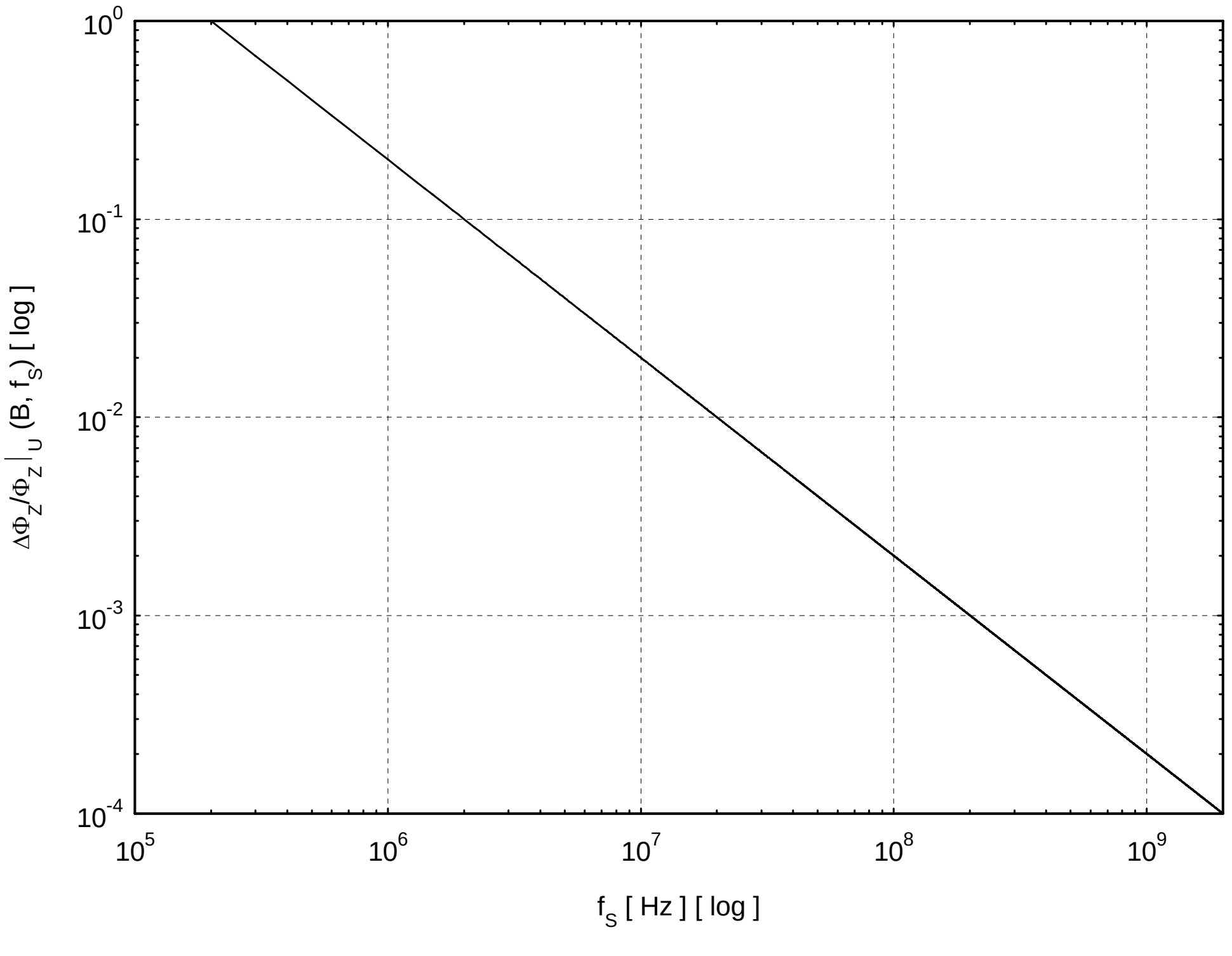

Figure 5.a

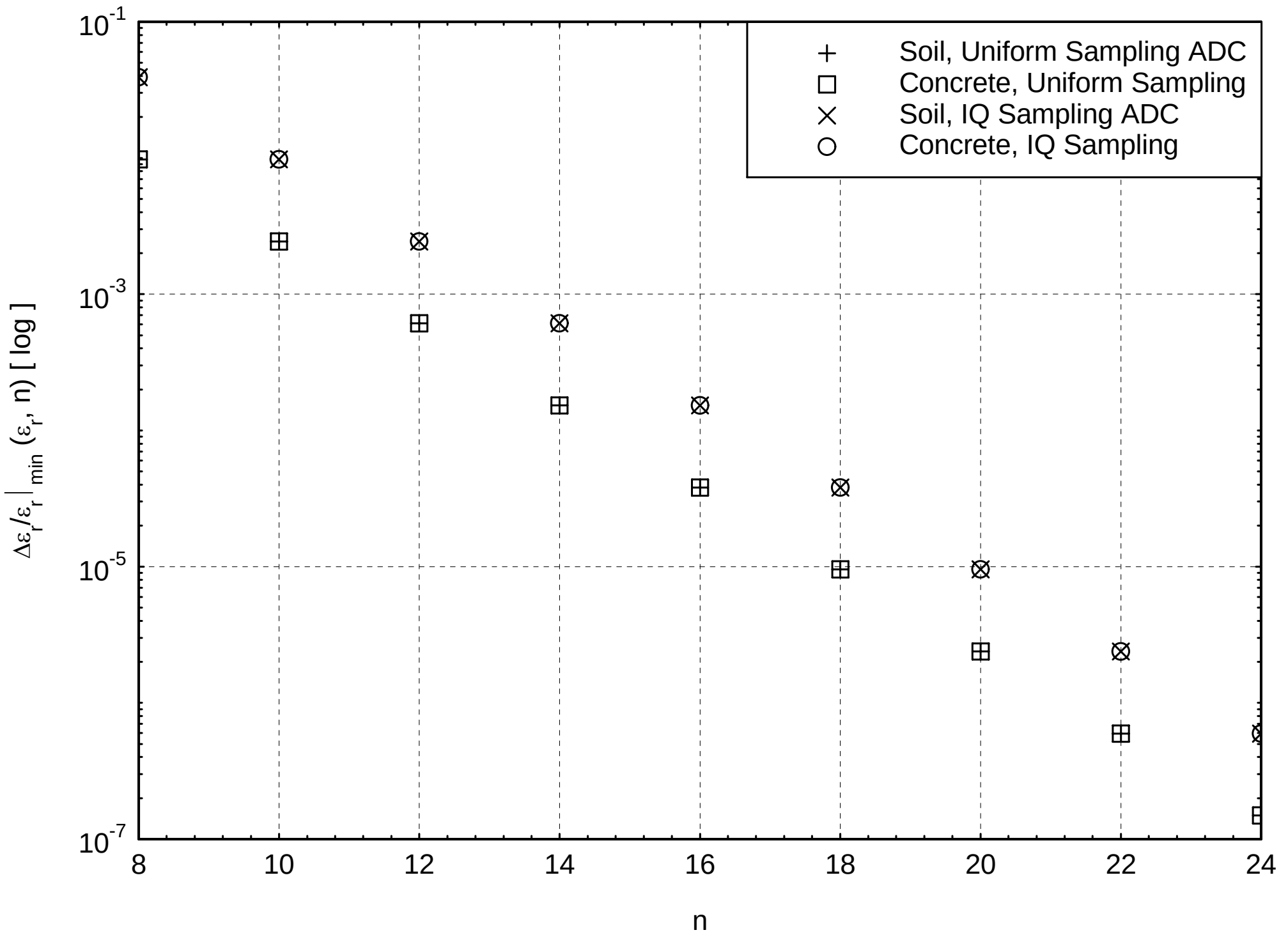

Figure 5.b

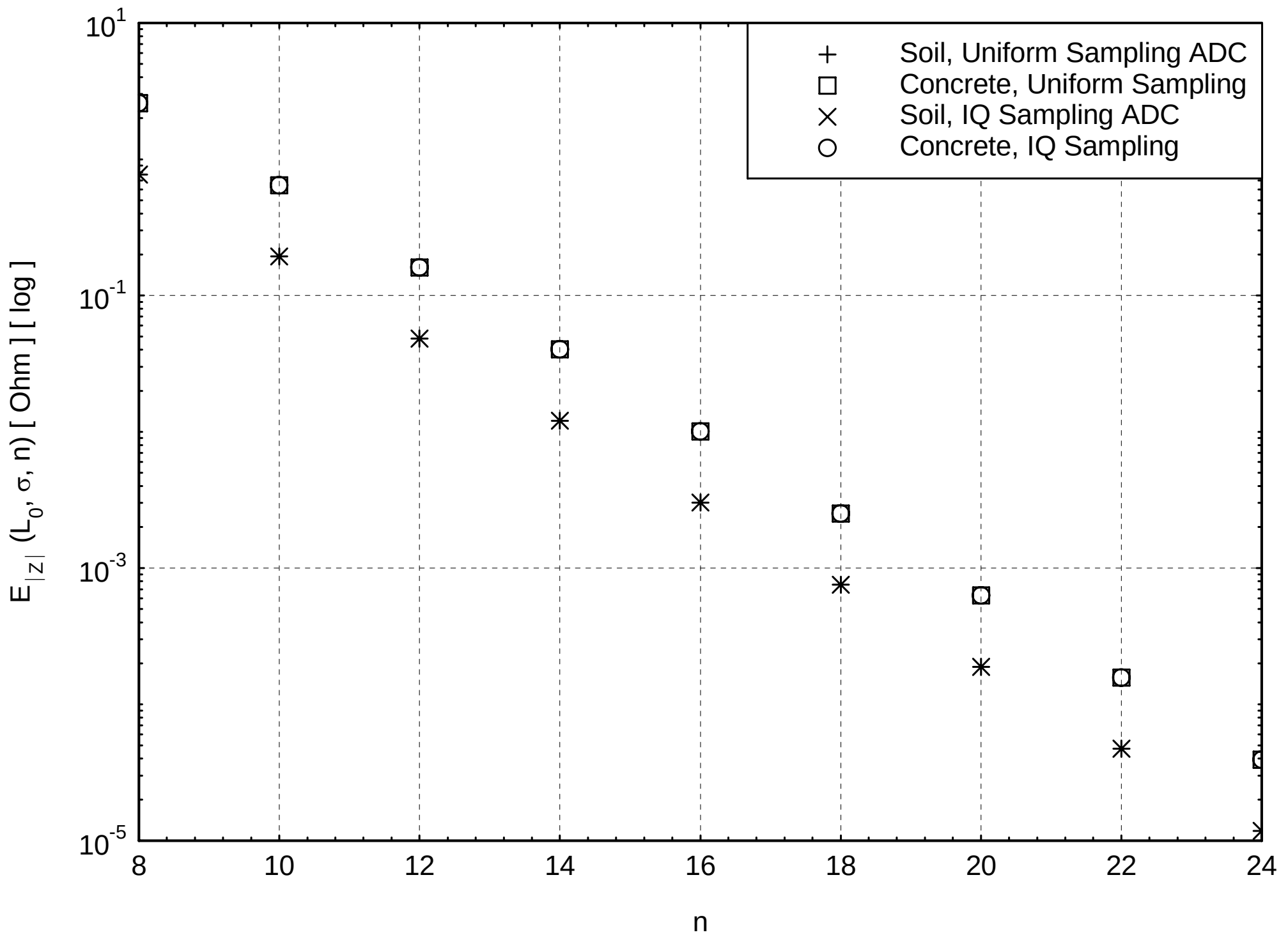

Figure 6.a

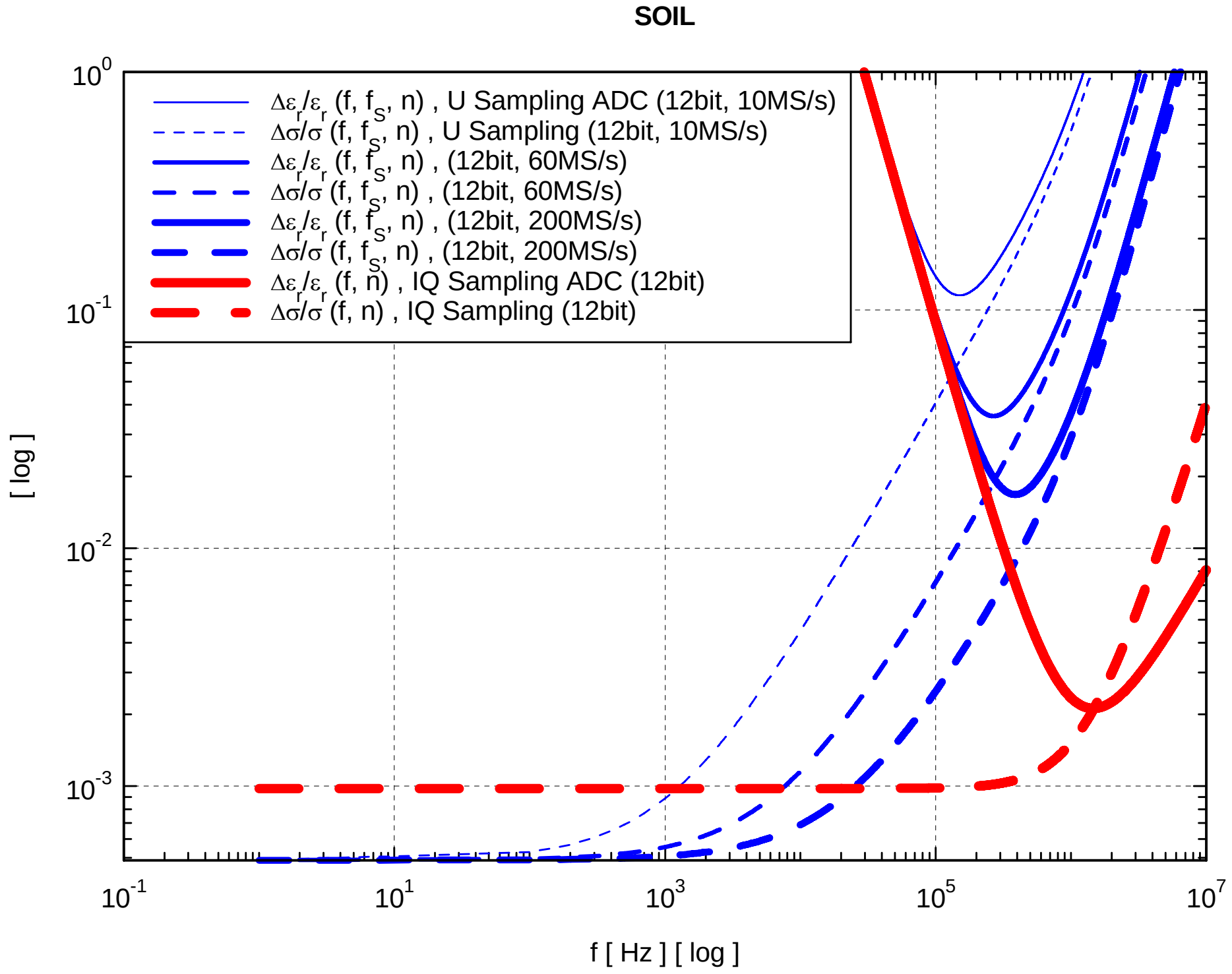

Figure 6.b

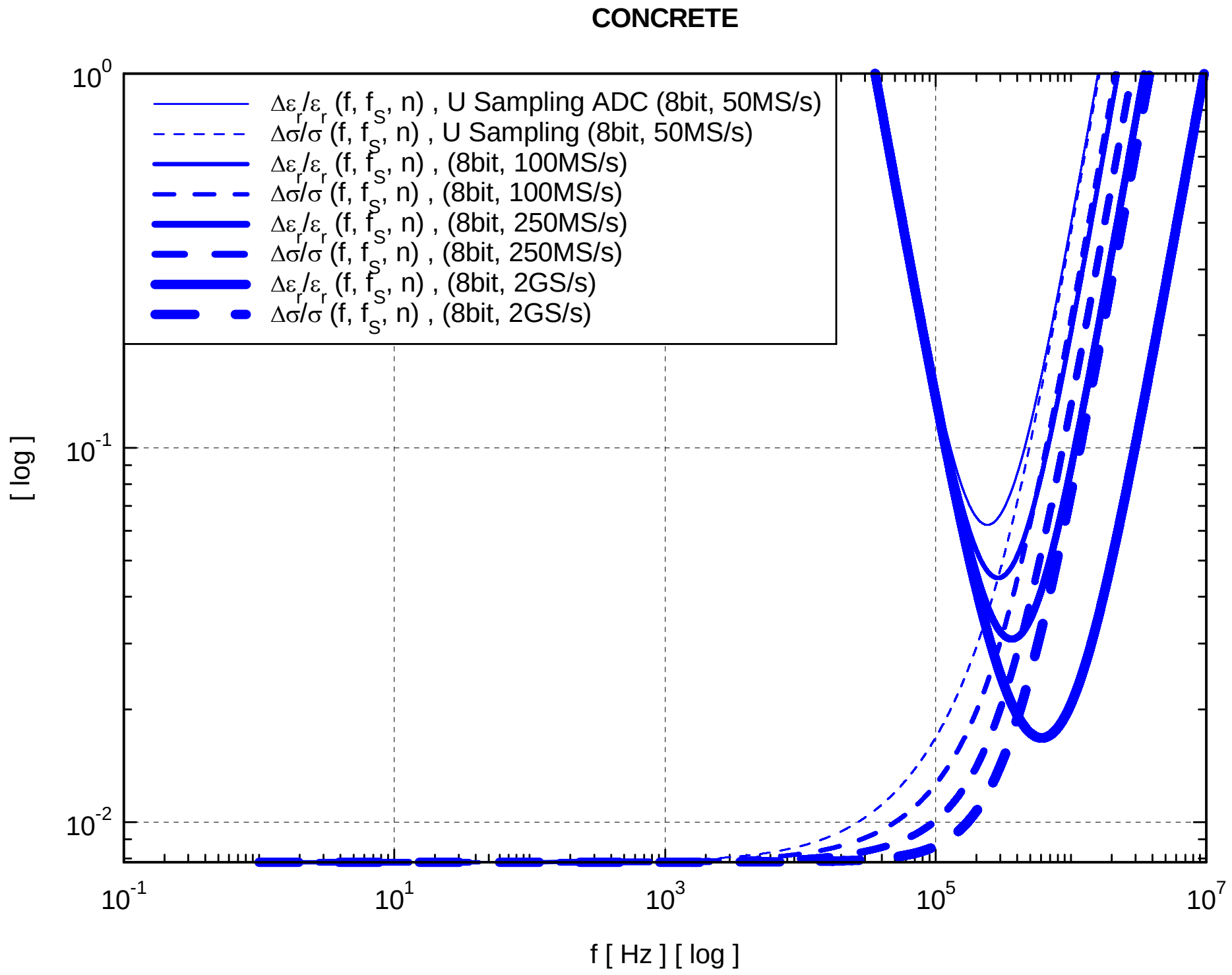

Figure 6.b.bis

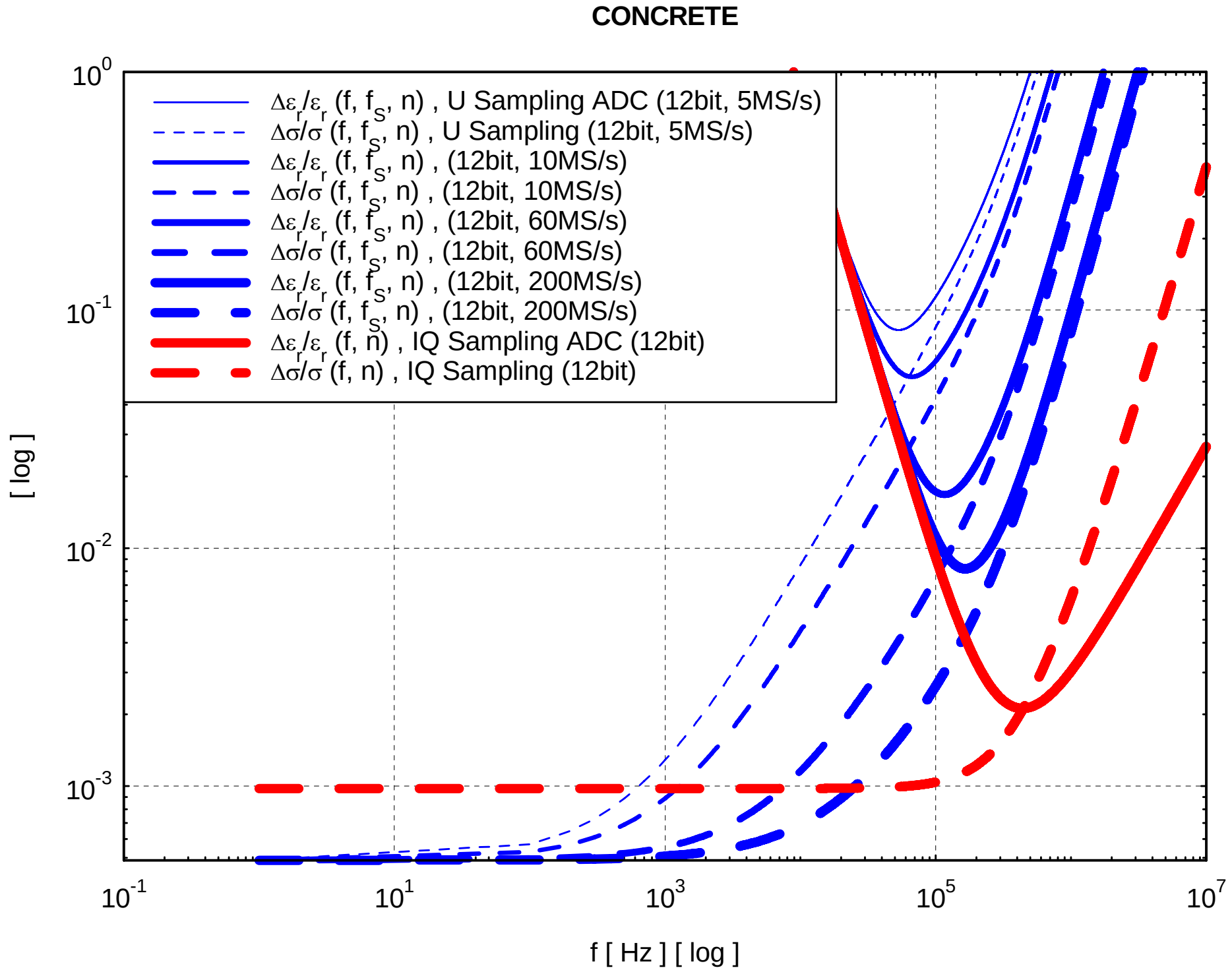

Figure 7.a

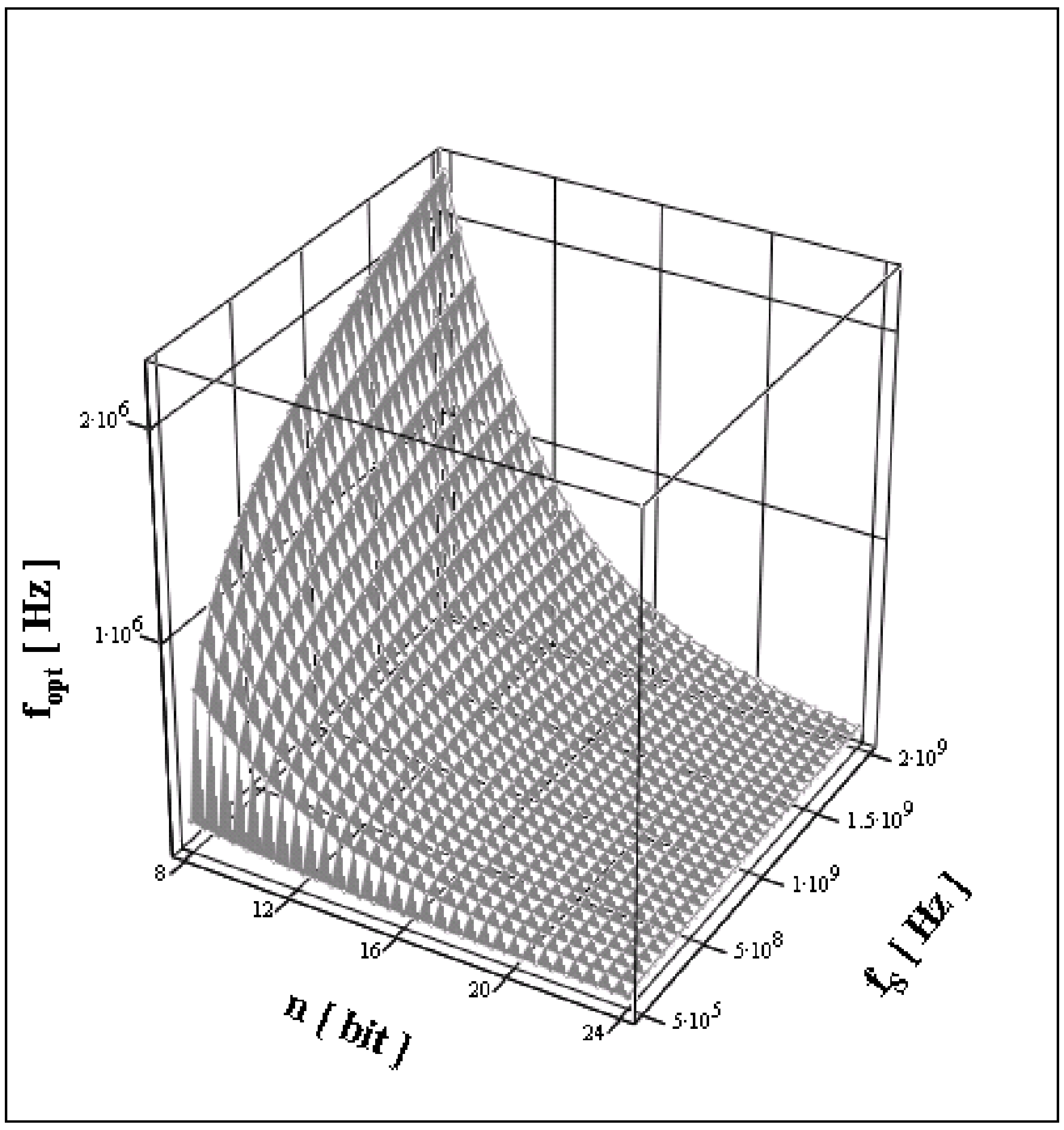

Figure 7.b

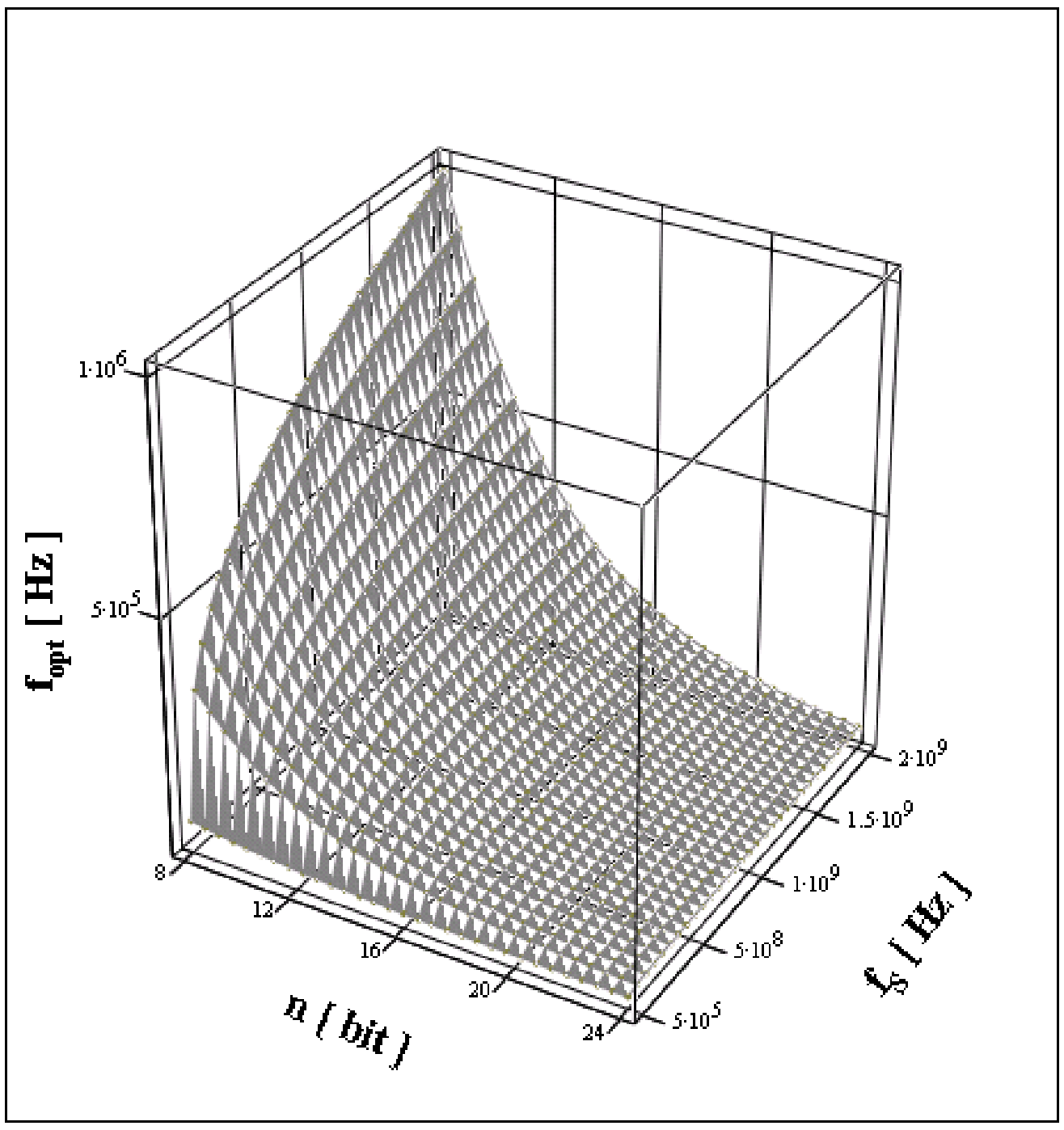

Figure 8

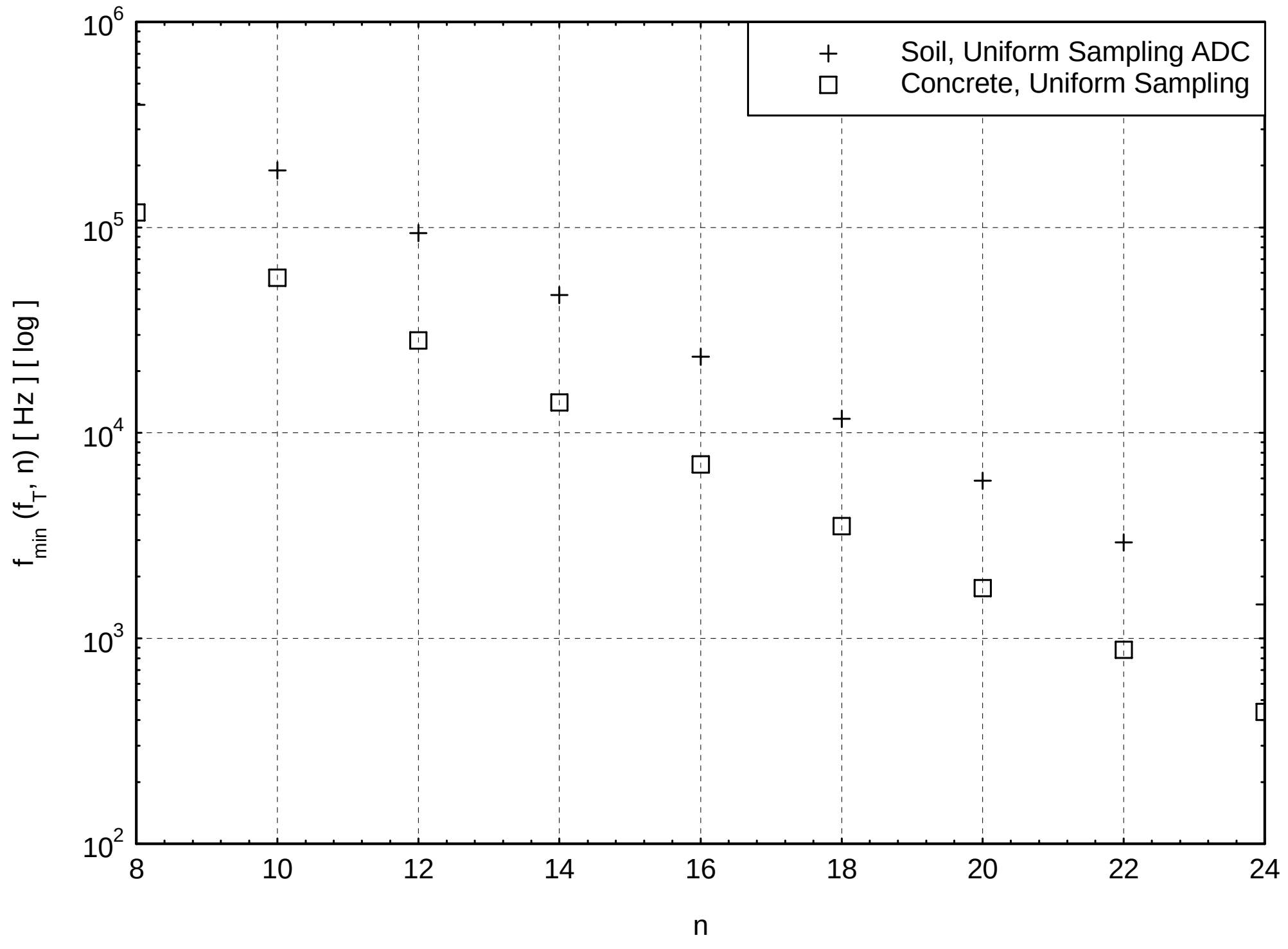

Soil, Uniform Sampling ADC
Concrete, Uniform Sampling
$f_{min}(f_T, n)$ [ Hz ] [ log ]
n
$10^6$
$10^5$
$10^4$
$10^3$
$10^2$
8
10
12
14
16
18
20
22
24

Figure 9.a

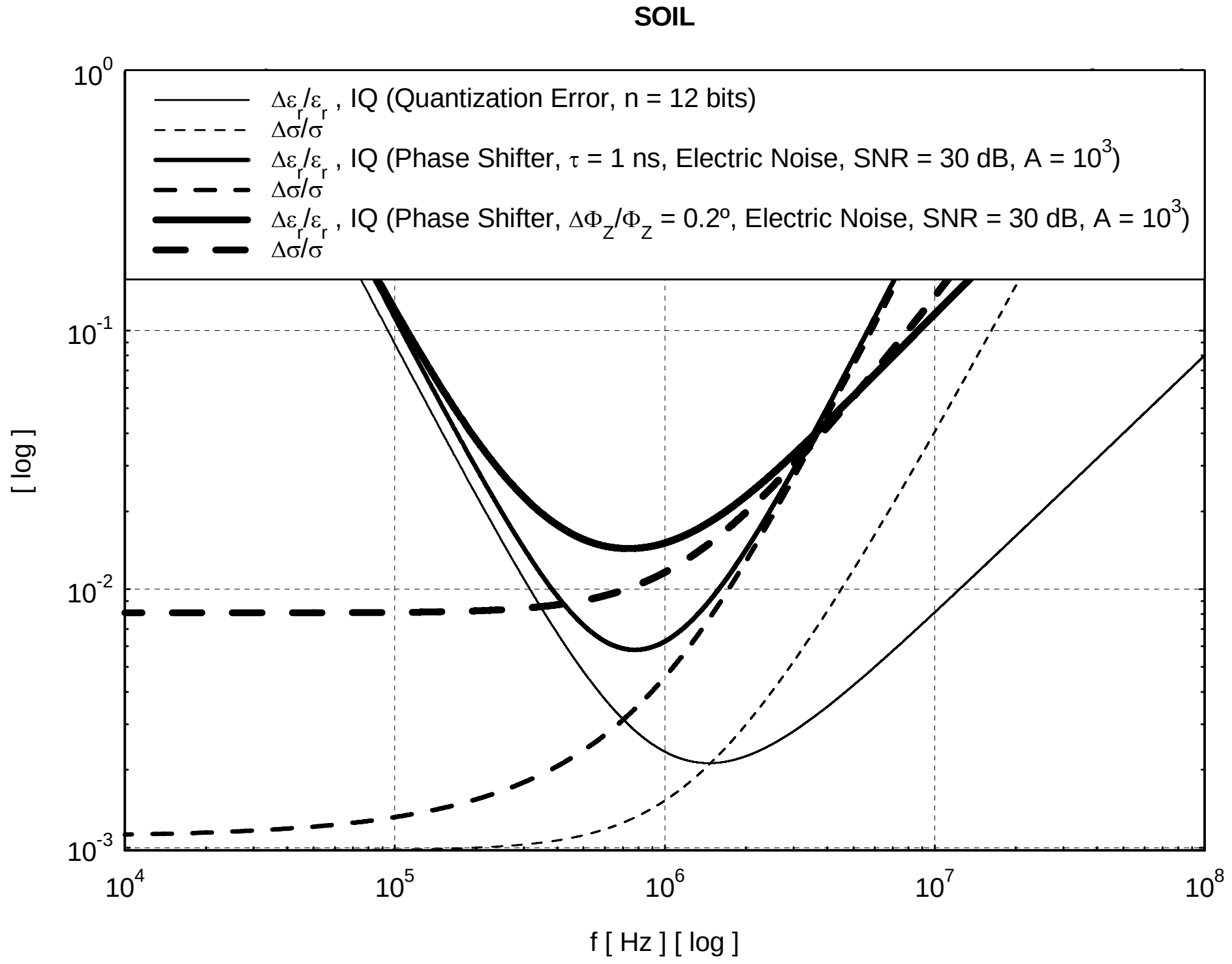

Figure 9.b

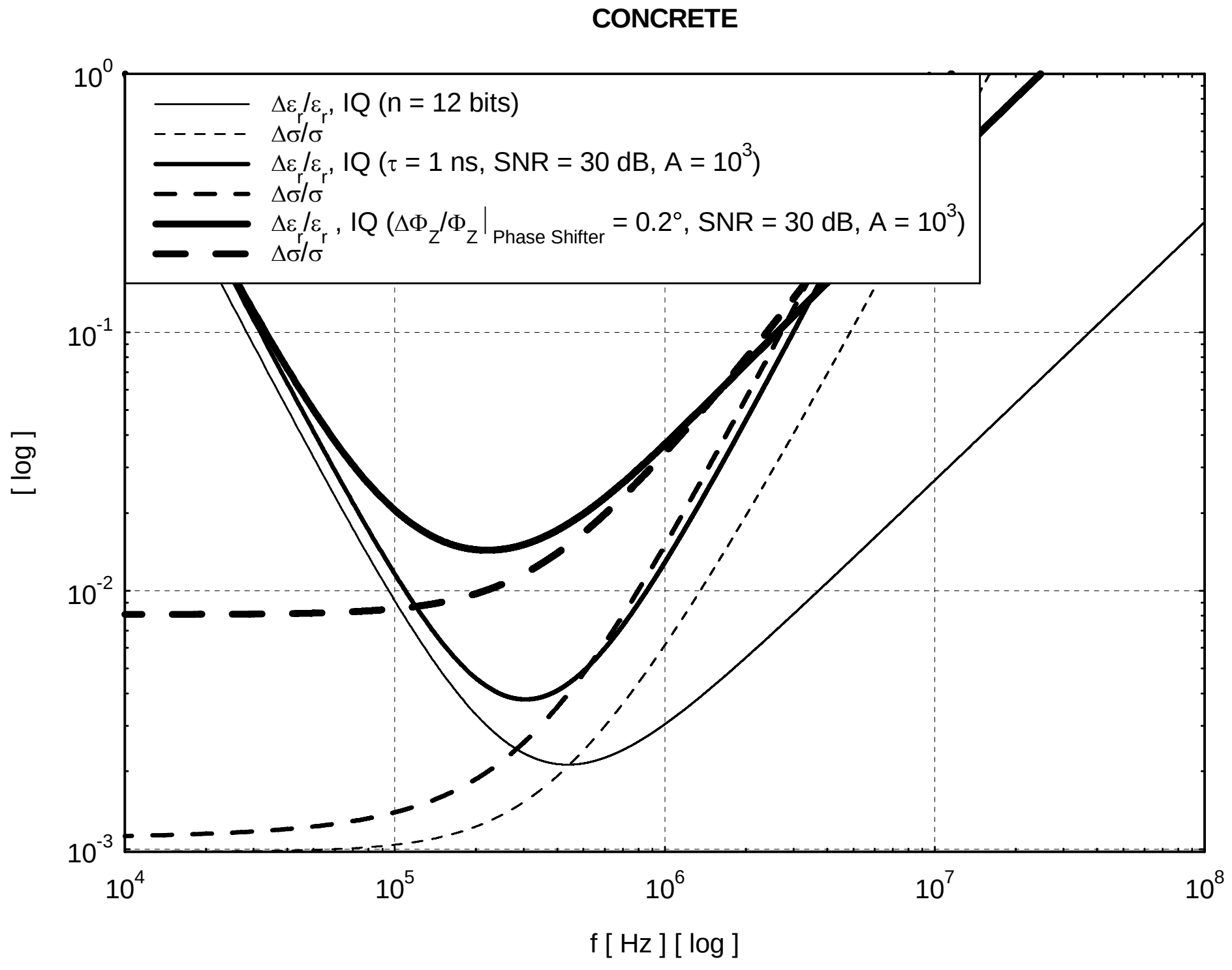

Figure 10.a

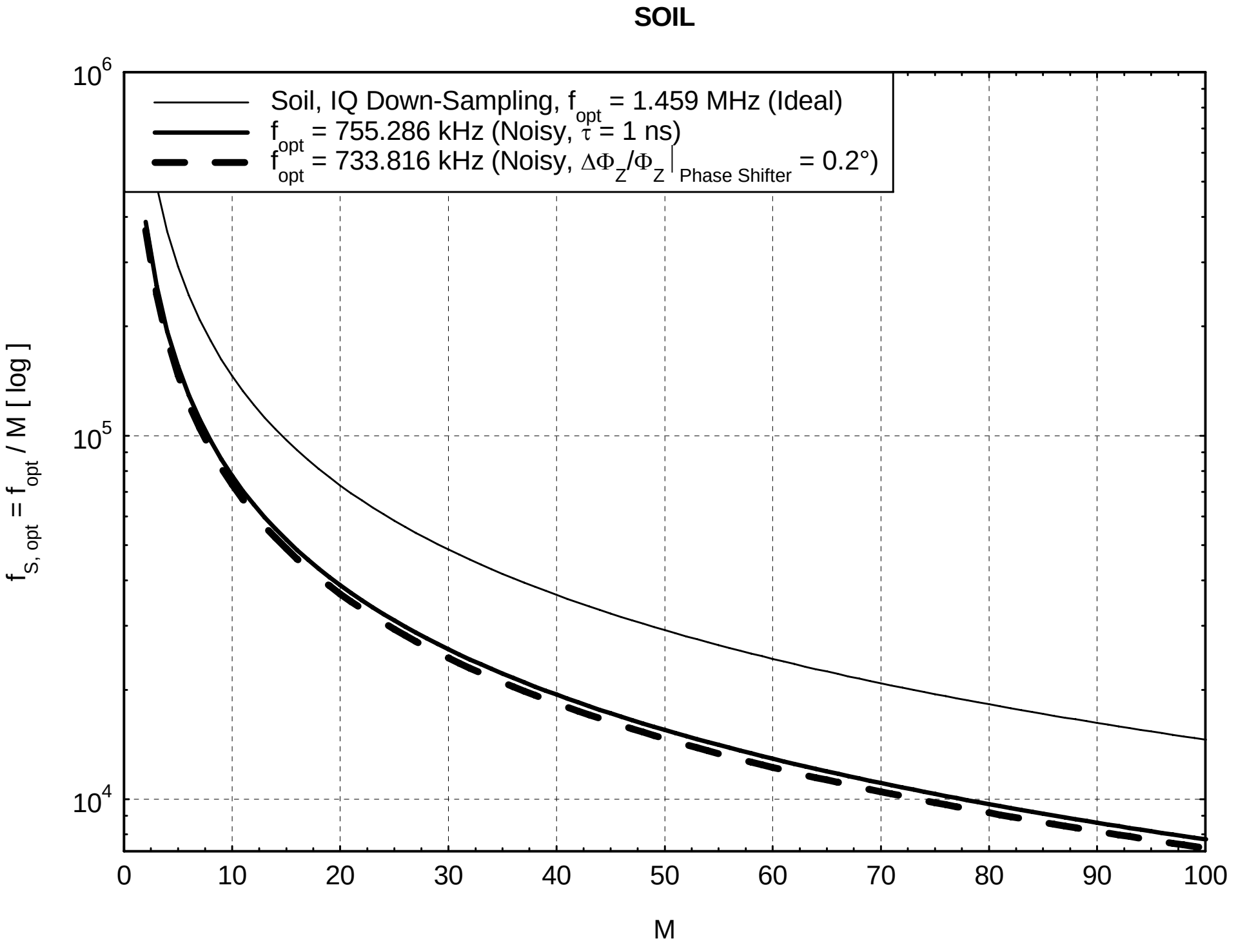

Figure 10.b

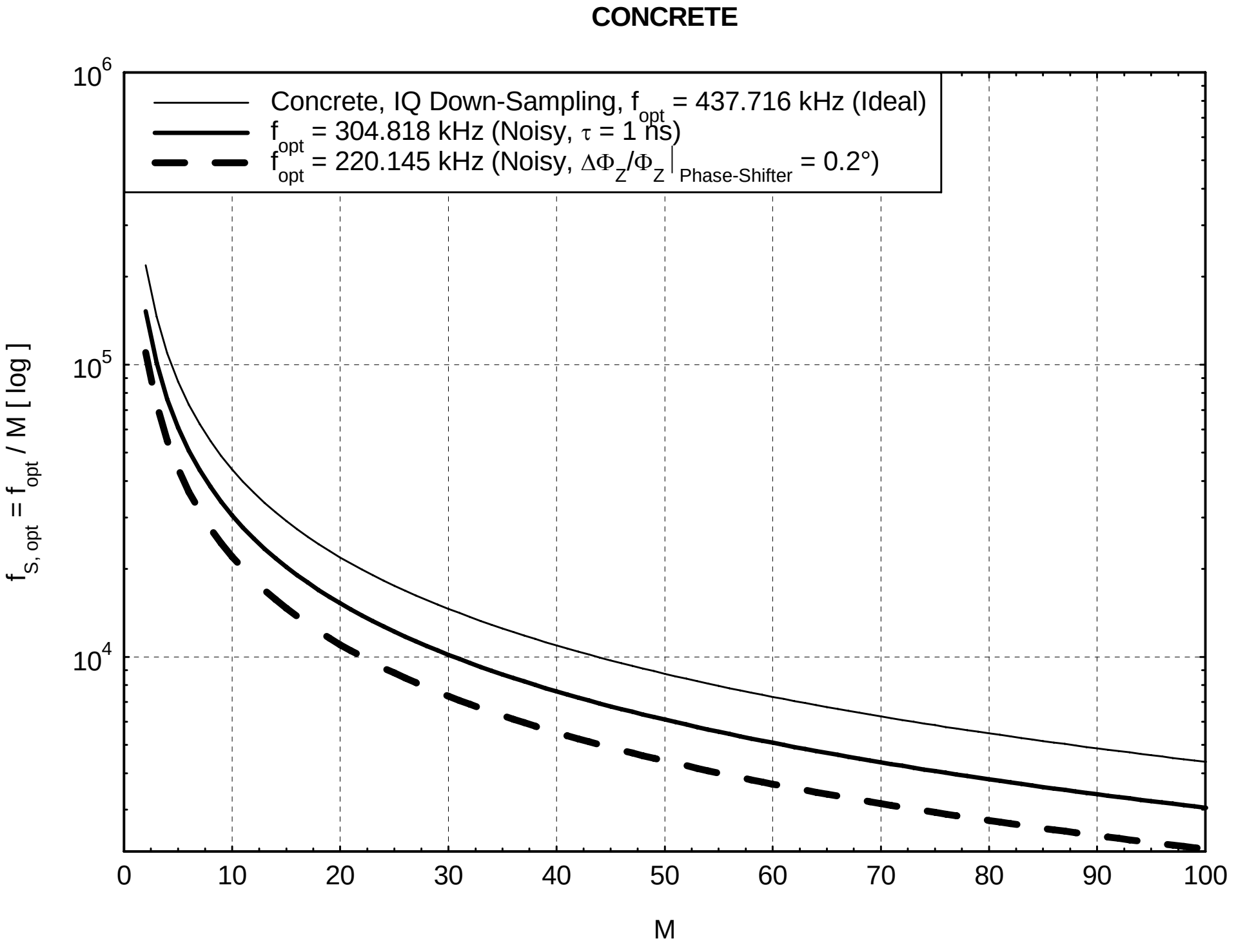

Figure 11.a

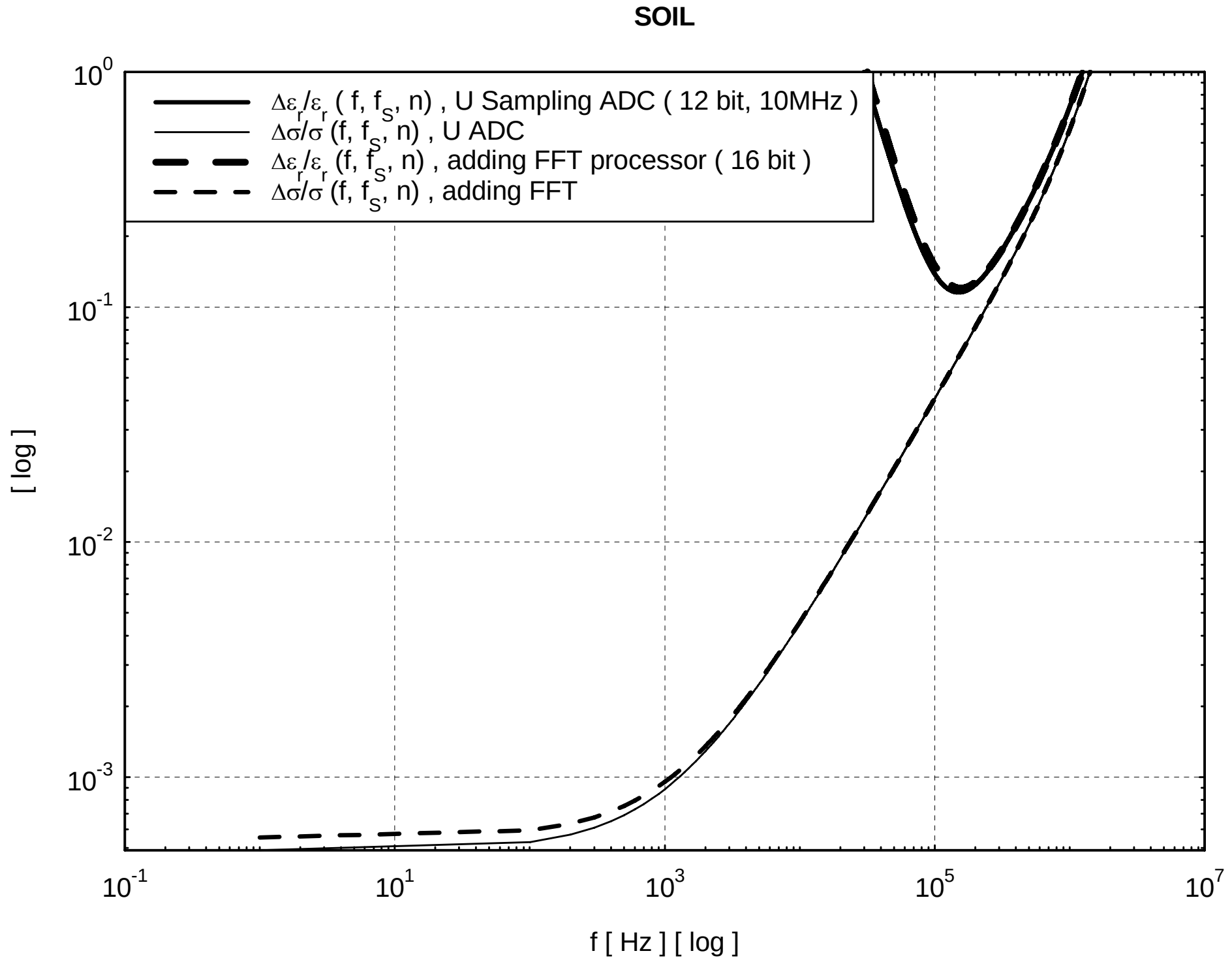

Figure 11.b

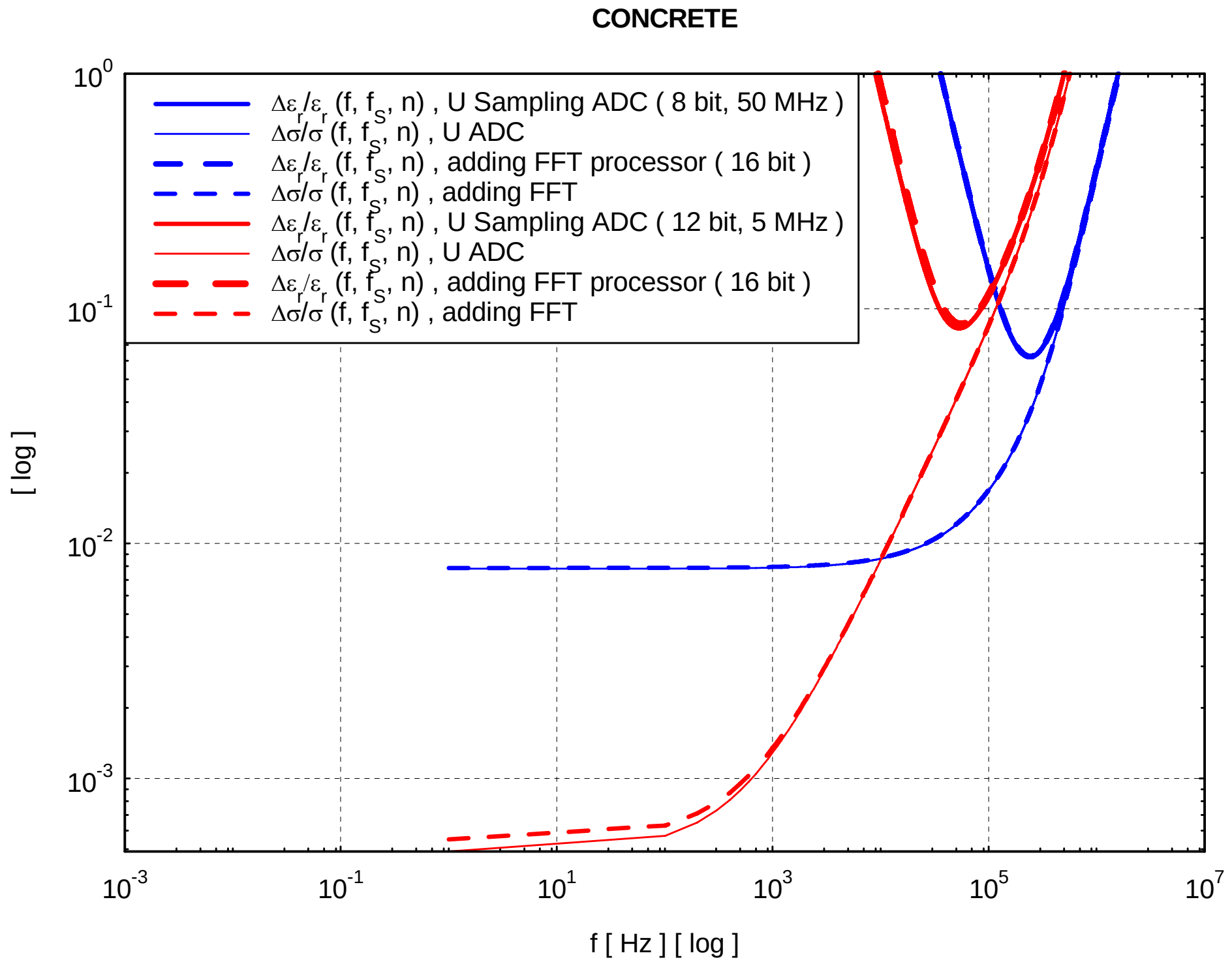

Figure 12.a

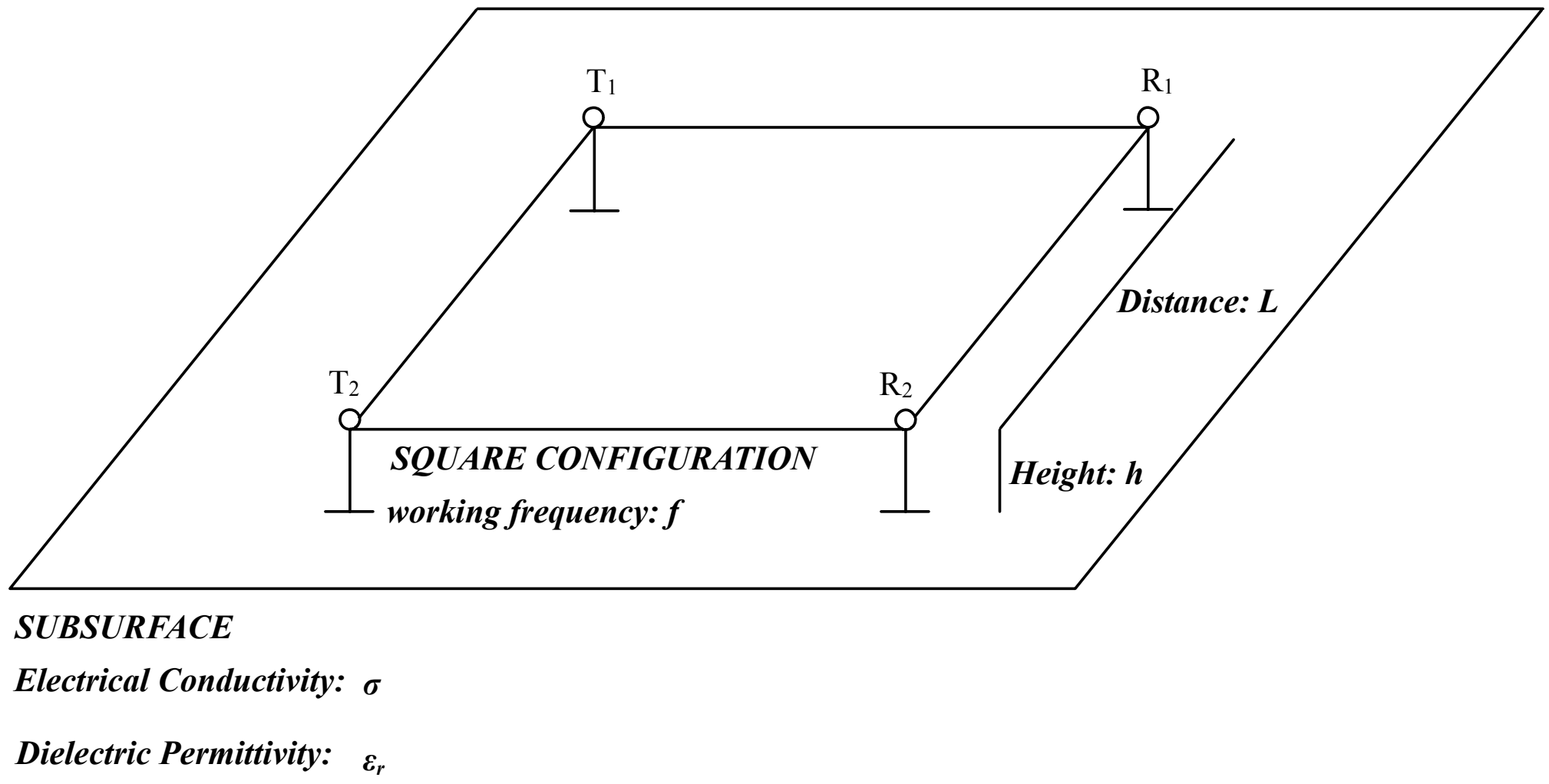

Figure 12.b

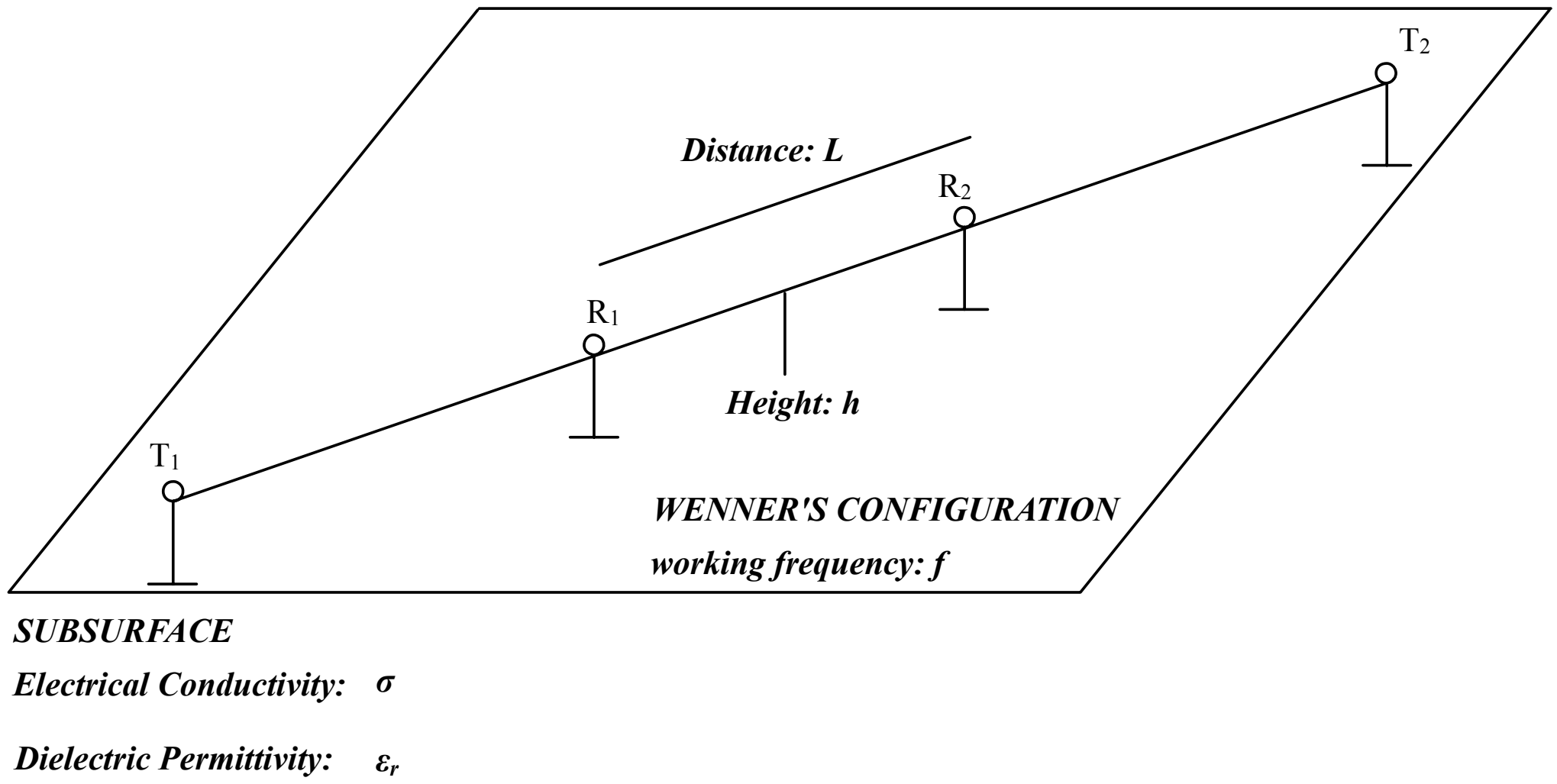

Figure 13.a

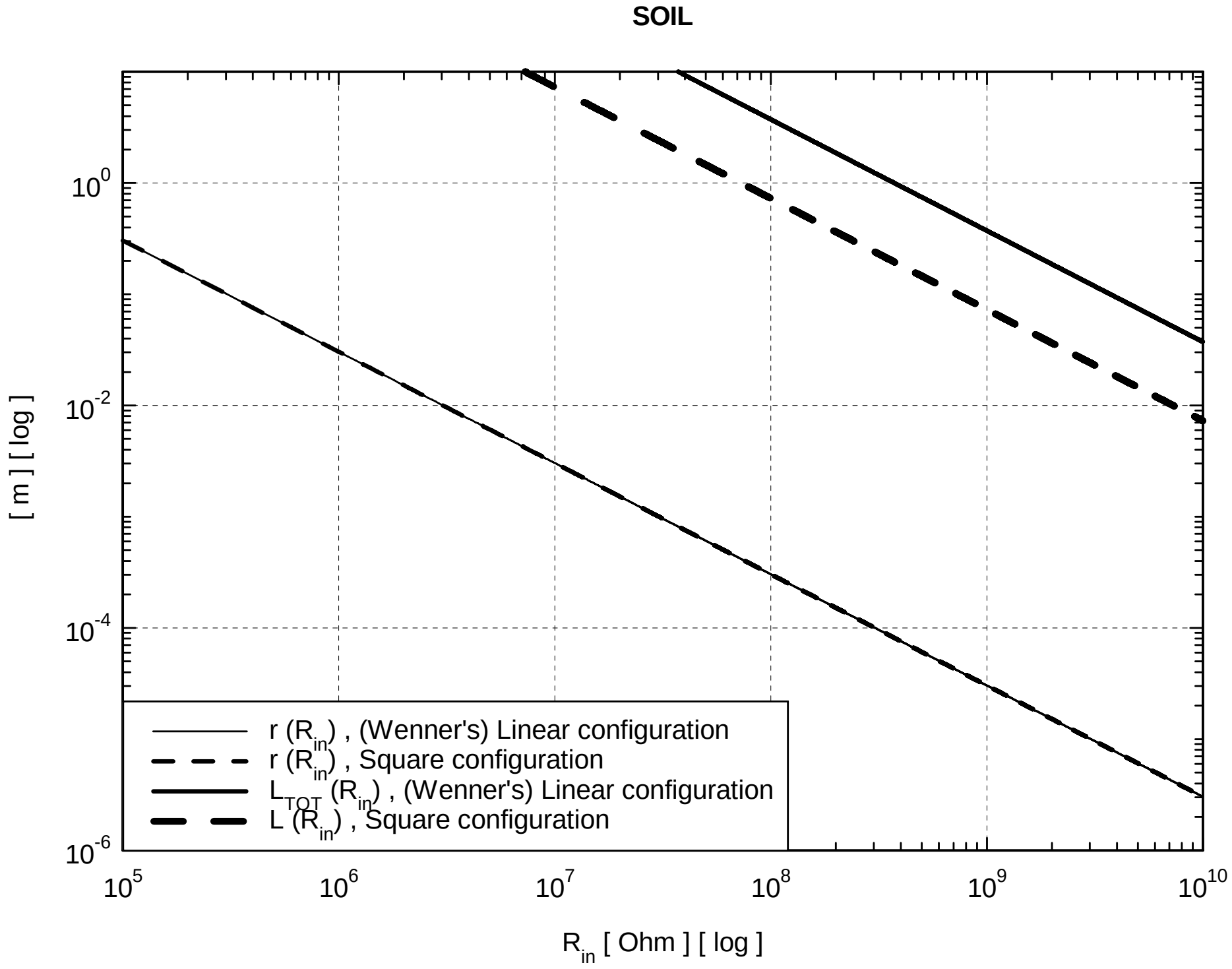

Figure 13.b

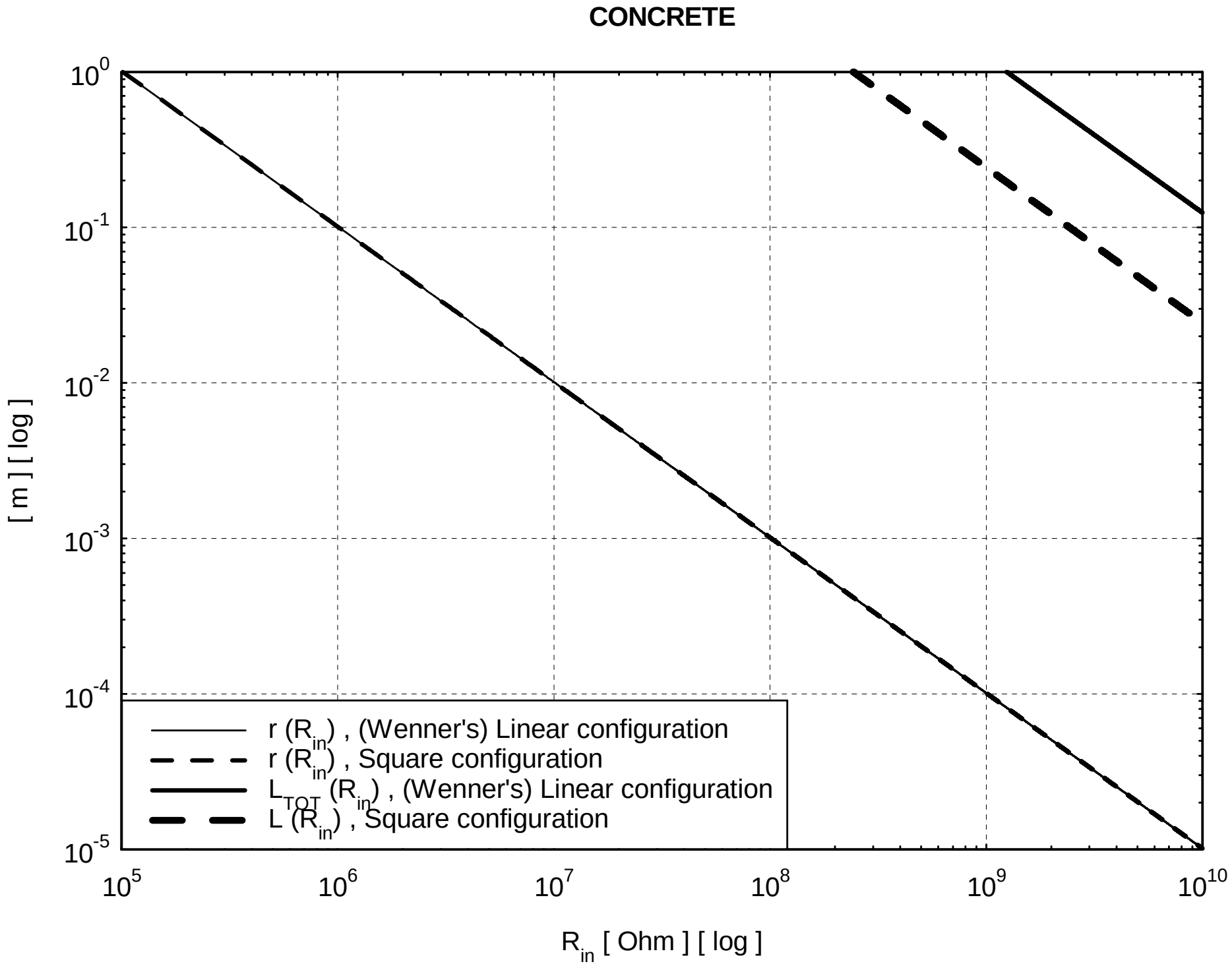

Figure 14.a

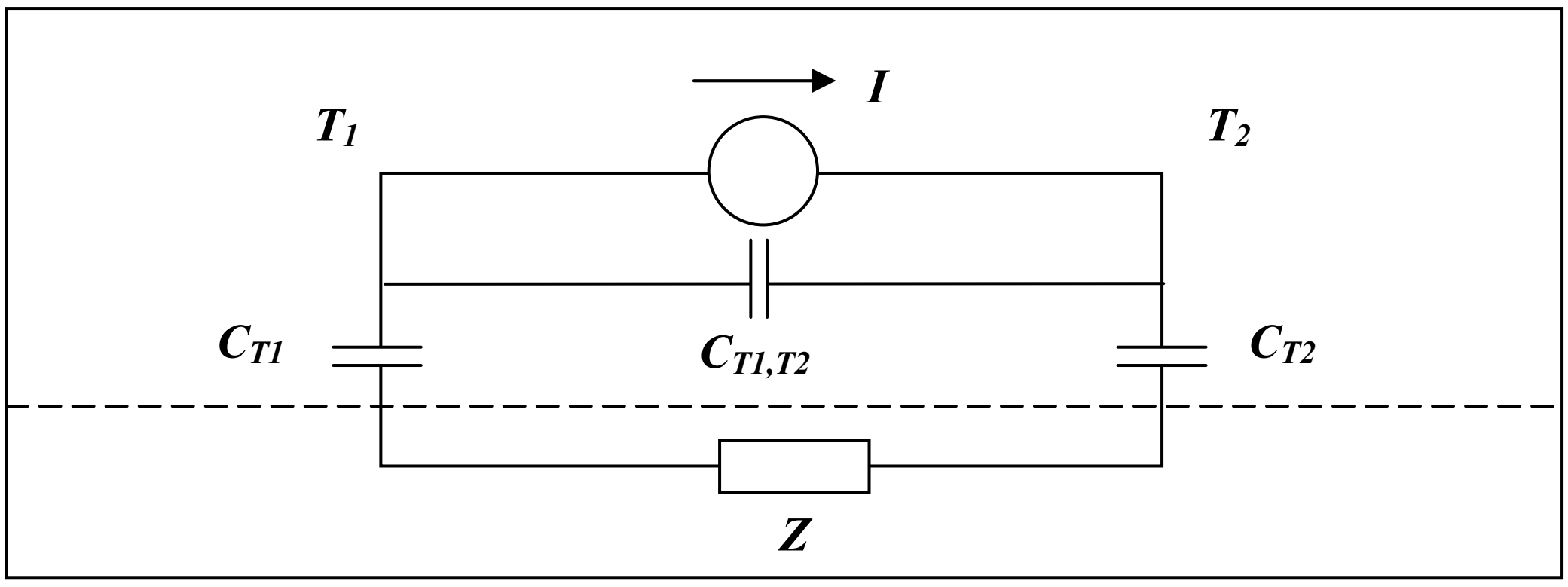

Figure 14.b

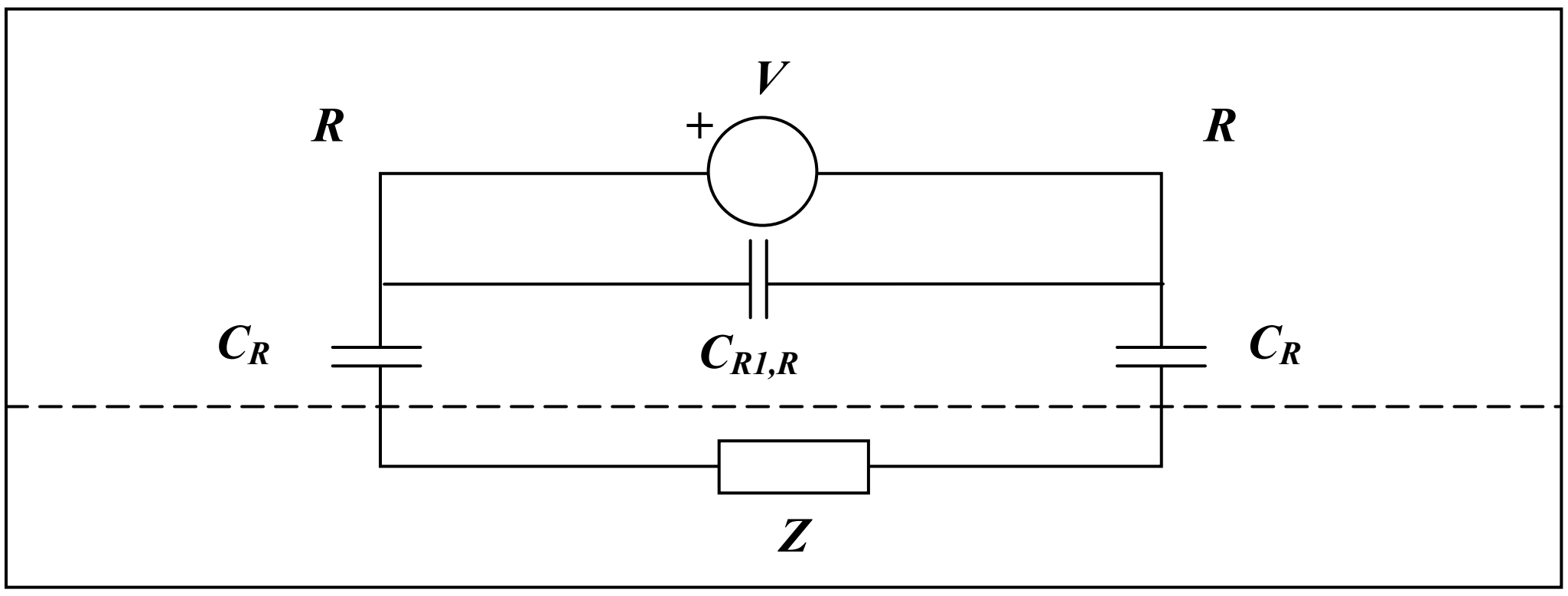

Fig.1.a. Equivalent circuit of the quadrupolar probe.

Fig. 1.b. Block diagram of the measuring system, which is composed of: a series of four electrodes laid on the material to be investigated; an analogical circuit for the detection of signals connected to a high voltage sinusoidal generator; a digital acquisition system; and a personal computer. Starting from the left, the four electrodes can be seen laid on the block of material to be analyzed. Two electrodes are used to generate and measure the injected current (at a selected frequency), while the other two electrodes are used to measure the potential difference. In this way, two voltages are obtained: the first proportional to the current; the second proportional to the difference of potential. These voltages are digitized through a digital analogical converter connected to a personal computer for further processing. The real magnitudes hereby measured in the time domain are subsequently transformed into complex magnitudes in the frequency domain. From the ratio of the complex values, at the specific investigated frequency, it is possible to obtain the complex impedance. A program with an algorithm of numerical inversion allows the electrical resistivity and dielectric permittivity of the material to be obtained by measuring the transfer impedance; in this way, the reliability of the measured data is immediately analyzed, proving very useful during a measurement program.

Fig 2.a. Practical scheme of an in phase and quadrature (IQ) down-sampling process. Samples I and Q can be enabled depending on whether the discrete time $n$ is even or odd ($n$ is a natural number). The sample in phase I is picked up at time $n \cdot M \cdot T$ and the sample in quadrature Q at time $n \cdot M \cdot T + T/4$ ($T$ is the signal period and $M$ the down-sampling factor). Obviously one can also choose a different values of $M$ shown by the figure ($M = 4$).

Fig 2.b. Logical scheme of a sampling and holding circuit (S&H) employing two IQ ADCs. The frequency $f$ of input signal is forwarded to a $90^{\circ}$ degree phase-shifter. Two chains of identical

programmable counters operate a division for the down-sampling factor $M$. The rate is precisely $f_S = f/M$.

Fig. 3. A quadrupolar probe, measuring an electrical voltage $V$ of initial phase $\varphi_V \in [0, 2\pi]$, is connected to an IQ sampler of bit resolution $n=12$, in which a quartz oscillation is characterized by a high figure of merit $Q \in [10^4, 10^6]$. Plots for the inaccuracies $\Delta|Z|/|Z|_{IQ}(\varphi_V, Q)$ and $\Delta\Phi_Z/\Phi_Z|_{IQ}(\varphi_V, Q)$ in the measurement of the modulus and phase for the transfer impedance: (a, b, semi-logarithmic) as functions of the phase $\varphi_V$, fixing the merit figure $Q$ within the range $[Q_{low}, \infty)$ from the lower limit $Q_{low} \approx 2\pi$, above which eq. (3.1) holds approximately; and (a.bis, b.bis, like-Bode's diagrams) as functions of $1/Q$, fixing $\varphi_V$ in the range $[\varphi_{V,min}, \varphi_{V,max}]$ from the minimum $\varphi_{V,min}$ till the maximum $\varphi_{V,\max}$ of eq. (3.1) (a.bis) or in the limit $\varphi_{V,lim}=\pi/2$ of eq. (3.6) (b.bis).

Fig. 4. A class of uniform ADCs is specified by bit resolution $n$, ranging from *8 bit* to *24 bit*, and rate sampling $f_S$, in the frequency band *[500 kHz, 2GHz]*: (a) semi-logarithmic plot for the inaccuracy $\Delta|Z|/|Z|_U(n)$ in the measurement of the modulus for transfer impedance, as a function of the resolution $n$; (b) Bode's diagram for the inaccuracy $\Delta\Phi_Z/\Phi_Z|_U(B, f_S)$ of the transfer impedance in phase, plotted as a function of the rate $f_S$, when the quadrupolar probe works in the upper frequency at the limit of its band $B=100kHz$.

Fig. 5. A quadrupole of characteristic geometrical dimension $L_0=1m$ exhibits a galvanic contact on a subsurface of dielectric permittivity around $\varepsilon_r \approx 4$ and low electrical conductivity $\sigma$, like a non-saturated terrestrial soil ($\sigma \approx 3.334 \cdot 10^{-4}$ $S/m$) or concrete ($\sigma \approx 10^{-4}$ $S/m$). In the hypothesis that the probe is connected to a sampler of bit resolution $n$, ranging from *8 bit* to *24 bit*: (a) semi-logarithmic plot for the "physical bound" imposed on the inaccuracies, i.e. $\Delta\varepsilon_r/\varepsilon_r|_{min}(\varepsilon_r, n)$, as a function of the resolution $n$; (b) semi-logarithmic plot for the absolute error $E_{|Z|}(L_0, \sigma, n)$ for the transfer impedance in modulus below its cut-off frequency, as a function of the resolution $n$.

Fig. 6. Refer to the operating conditions described in the caption of fig. 5 (for quadrupole and subsurfaces). Bode's diagrams for the inaccuracies $\Delta\varepsilon_r/\varepsilon_r(f)$ and $\Delta\sigma/\sigma(f)$ in the measurement of the dielectric permittivity $\varepsilon_r$ and the electrical conductivity $\sigma$, plotted as functions of the frequency $f$, for non-saturated terrestrial soil (a) or concrete (b, c) analysis. The probe is connected to uniform or IQ or samplers (in the worst case, when the internal quartz is oscillating at its lowest merit factor $Q \approx 10^4$), of bit resolution $n=12$ (a, c) or $n=8$ (b), which allow inaccuracies in the measurements below a prefixed limit, *15%* referring to (a, b) and *10%* for (a, c), within the frequency band $B=100kHz$ [Tab. 1].

Fig. 7. Refer to the operating conditions described in the caption of fig. 4 (for uniform sampling ADCs) and fig. 5 (for quadrupolar probe). Plots for the optimal working frequency $f_{opt,U}(n,f_S)$, which minimizes the inaccuracy in the measurement of dielectric permittivity, as a function of both the bit resolution $n$ and the sampling rate $f_S$, for terrestrial soil (a) or concrete (b) analysis.

Fig. 8. Refer to the operating conditions described in the caption of fig. 5 (subsurface: non-saturated terrestrial soil or concrete; sampling: uniform). Plot for the minimum value of frequency $f_{U,min}(n,f_T)$, which allows an inaccuracy in the measurement of the dielectric permittivity below a prefixed limit (*10%*), as a function of the bit resolution $n$ [the cut-off frequency of the transfer impedance fixed as $f_T=f_T(\varepsilon_r,\sigma)$ corresponding to the subsurface defined by the measurements $(\varepsilon_r,\sigma)$].

Fig. 9. Refer to the caption of fig. 6 [terrestrial soil (a) or concrete (b) analysis]. Here, the quadrupole is connected to an IQ sampler of bit resolution $n = 12$ and it is affected by an additional noise both in amplitude, due to the external environment (signal to noise ratio $SNR = 30\ dB$, averaged terms $A = 10^3$), and in phase, due to a phase-splitter detector specified by (rise time, $\tau = 1ns$) or (phase inaccuracy, $\Delta\Phi_Z/\Phi_Z = 0.2°$). The noisy probe allows inaccuracies $\Delta\varepsilon_r/\varepsilon_r(f)$ and $\Delta\sigma/\sigma(f)$ in the measurement of permittivity $\varepsilon_r$ and conductivity $\sigma$ below a prefixed limit, *15%* referring to (a) and *10%* for (b), within the frequency band $B=100kHz$ [Tab. 2].

Fig. 10. Refer to the caption of fig. 9 [terrestrial soil (a) or concrete (b) analysis]. An ideal or noisy quadrupole works at the optimal frequency $f_{opt}$ which minimizes the inaccuracy $\Delta\varepsilon_r/\varepsilon_r(f)$ in measurement of permittivity $\varepsilon_r$ [see Tab. 2]. The IQ technique can be performed at a rate $f_{S,opt}(M) = f_{opt}/M$ inversely proportional to the down-sampling factor $M$.

Fig. 11. Refer to the caption of fig. 6 [terrestrial soil (a) or concrete (b) analysis]. Here, the quadrupolar probe is connected to an uniform ADC of minimal bit resolution $n\leq12$ and over-sampling rate $f_S$, (*12 bit, 10 MHz*)(a) and (*8 bit, 50 MHz*) or (*12 bit, 5 MHz*)(b), in addition to a FFT processor of register length $n_{FFT} = 16$ which allow inaccuracies $\Delta\varepsilon_r/\varepsilon_r(f)$ and $\Delta\sigma/\sigma(f)$ in the measurement of permittivity $\varepsilon_r$ and conductivity $\sigma$ below a prefixed limit, *15%* referring to (a) and *10%* for (b), within the frequency band $B=100kHz$ [Tab. 3].

Fig. 12. Quadrupolar probe in square (a) or linear (Wenner's) (b) configuration.

Fig. 13. Refer to the operating conditions described in the caption of fig. 3 (for quadrupolar probe); the quadrupole is connected to an IQ sampler of minimum bit resolution $n_{min}=12$, which allows inaccuracies in the measurements below a prefixed limit (*10%*) within the band $[f_{min}, B]$ from the minimum value of frequency $f_{min}$ to $B=100kHz$. Like-Bode's diagrams of the radius $r(R_{in},f_{min})$ for probe electrodes and of the characteristic geometrical dimension $L_S(r,n_{min})$ for square or of the length $L_{TOT}(r,n_{min})=3\cdot L_W(r,n_{min})$ for (Wenner's) linear configurations, plotted as functions of the input resistance $R_{in}$ for the amplifier stage [non saturated terrestrial soil (a) and concrete (b)][Tab. 4].

Fig. 14. Equivalent capacitance circuits of the quadrupolar probe which schematize the transmission (a) and reception (b) stage.